\documentclass[showpacs,prd]{revtex4}

\usepackage{epsfig}
\usepackage{graphicx}
\usepackage{dcolumn}
\usepackage{amsmath}
\usepackage{latexsym}

\newcommand{\pder}[2]{\frac{\partial#1}{\partial#2}}
\newcommand{\re}[1]{(\ref{eq:#1})}
\newcommand{\bra}[1]{\left\langle#1\phantom{.}\!\right|}
\newcommand{\ket}[1]{\left|\phantom{.}\!#1\right\rangle}
\newcommand{\psih}{\hat{\psi}}
\newcommand{\no}[1]{:#1\!\,:}
\newcommand{\mycomment}[1]{}

\newcommand{\ltot}{L_3^{\rm (tot)}}
\newcommand{\nmax}{N^{\rm (max)}}

\begin{document}

\title{The thermodynamics of Fermi gases in three dimensional fuzzy space}

\author{F~G~Scholtz$^{1,2}$, J~N~Kriel$^1$ and H~W~Groenewald$^1$}
\address{$^1$Institute of Theoretical Physics, University of Stellenbosch, Stellenbosch 7600, South Africa}
\address{$^2$National Institute for Theoretical Physics (NITheP), Stellenbosch 7600, South Africa}

\begin{abstract}
We use the recently derived density of states for a particle confined to a spherical well in three dimensional fuzzy space to compute the thermodynamics of a gas of non-interacting fermions confined to such a well.  Special emphasis is placed on non-commutative effects and in particular non-commutative corrections to the thermodynamics at low densities and temperatures are computed where the non-relativistic approximation used here is valid.  Non-commutative effects at high densities are also identified, the most prominent being the existence of a minimal volume at which the gas becomes incompressible.  The latter is closely related to a low/high density duality exhibited by these systems, which in turn is a manifestation of an infra-red/ultra violet duality in the single particle spectrum.  Both  non-rotating and slowly rotating gasses are studied.  Approximations are benchmarked against exact numerical computations for the non-rotating case and several other properties of the gas are demonstrated with numerical computations. Finally, a non-commutative gas confined by gravity is studied and several novel features regarding the mass-radius relation, density and entropy are highlighted.
\end{abstract}
\vspace{2pc}

\maketitle

\section{Introduction}

The structure of space-time at short length scales is probably the most challenging problem facing modern physics \cite{seib}.  Although there seems to be growing consensus that our notion of space-time at the Planck length needs major revision, there seems to be much less consensus on the appropriate description of space-time at these length scales.  One scenario, originally proposed by Snyder \cite{Snyder},  is that of non-commutative space-time. Although his motivation for this proposal was an attempt to avoid the infinities encountered in quantum field theories, the notion of non-commutative space-time has recently been revived by the rather compelling arguments of Doplicher et al \cite{dop} and the emergence of non-commutative coordinates in the low energy limit of certain string theories \cite{Seiberg}.

Since the energies required to probe the Planck length scale only occurred in the very early universe, we do not have any direct empirical data to guide our search for the structure of space-time at the Planck length scale.  Possible sources of indirect data are the cosmic micro-wave background radiation (CMB), other cosmological phenomena such as dark matter and dark energy, high energy cosmic radiation and cold, dense astro-physical objects such as neutron stars and white dwarfs.  In particular it has been observed before \cite{nc-entropy} that the thermodynamic behaviour of two dimensional Fermi gases is drastically altered at high densities and low temperatures.  As these computations were performed for two dimensional systems, not much can be inferred about the behaviour of such real physical objects.  

This motivates the current paper in which the thermodynamic behaviour of Fermi gases in three dimensional non-commutative space is investigated.  To avoid the difficulties related to the breaking of rotational symmetry when the simplest constant commutation relations are assumed for the coordinates \cite{scholtz}, we rather adopt here fuzzy sphere commutation relations for the coordinates.  The recent solution for the spectrum of the non-commutative three dimensional fuzzy well \cite{nc-well} is then used to compute the thermodynamics of these gases, confined to a finite volume, in three dimensions.  Particular attention is paid to the behaviour of these gases under conditions where non-commutative effects are expected to be relevant, i.e. at high temperatures and in dense, cold Fermi gases.  These computations, based on solving the non-commutative Schr{\"o}dinger equation, are still non-relativistic and certainly not realistic for the ultra-relativistic conditions prevailing at high temperatures and in dense astro-physical objects.  However, although the non-relativistic description may affect the quantitative behaviour, one does not expect it to alter the qualitative behaviour, such as incompressibility.  We therefore expect to learn a considerable amount about the qualitative behaviour of such systems in a non-relativistic description, which is considerably simpler than the corresponding more realistic, but also much more difficult, relativistic description, which is as yet still lacking.   We do, however, also compute the non-commutative corrections at low densities and temperatures where the non-relativistic approximation is certainly appropriate. These corrections may offer useful guidance to attempts to observe these phenomena at low energies and densities.

In this paper the focus falls on Fermi gases, where one expects non-commutative effects to be most prominent.  One can, of course, also study the thermodynamics of Bose gases, but as one expects non-commutative phenomena to be less prominent there, we postpone this study to a future paper. Furthermore, at low densities these gases are essentially classical or Boltzmann, where quantum and statistical effects are unimportant, and the non-commutative corrections computed in this limit should show up regardless of whether we consider a Bose or Fermi gas. 

The paper is organised as follows:  Section \ref{Section 1} reviews the basic formulation of quantum mechanics in fuzzy space and the solutions of the infinite fuzzy well.  Section \ref{Section 2} sets up the formalism to study the thermodynamics of the non-commutative Fermi gas. In particular it is demonstrated that a remarkable infra-red/ultraviolet duality in the spectrum results in a similar duality between the low and high density limits of a non-commutative Fermi gas. Section \ref{Section 3} focuses on the computation of the central thermodynamic quantities, firstly, in the low density, low temperature as well as low density, high temperature limits.  The low/high density duality is then applied to infer from this the behaviour of these quantities in the high density limit.  Both non-rotating and slowly rotating gases are discussed.   Section \ref{Section 4} discusses a Fermi gas confined by gravity and in particular the effect of non-commutativity on the mass-radius relation, density and entropy.  Section  \ref{Section 5} summarises our results and draws conclusions.  

\section{Spectrum of the infinite fuzzy well}
\label{Section 1}

\subsection{Basic formalism}
We begin by summarizing the formalism of quantum mechanics in non-commutative three-dimensional space, as set out in \cite{nc-well,nc-coulomb}. The starting point is to impose on the spatial coordinates the $su(2)$ commutation relations 
\begin{equation}
	[\hat{X}_i,\hat{X}_j]=2i\theta \varepsilon_{ijk}\hat{X}_k
\label{eq:nc-coor-coms}
\end{equation}
where $\theta$ is the non-commutative length parameter.  The Casimir operator $\hat{X}^2=\hat{X}_i\hat{X}_i$ is naturally associated with the radial distance squared and its eigenvalues are determined by the $su(2)$ representation under consideration.

The first step in setting up the non-commutative quantum system is to build a representation of the non-commutative coordinate algebra (\ref{eq:nc-coor-coms}) on a Hilbert space $\cal{H}_C$, referred to as the configuration space.  In this case we want to mimic $R^3$ as a collection of spheres (an onion structure) where the squared radii of the spheres are quantised according to the eigenvalues of the $su(2)$ Casimir. For this $\mathcal{H}_C$ must carry a single copy of each su(2) irrep. A convenient, concrete realisation of this configuration space is provided by the Schwinger construction, which utilises two sets of boson creation and annihilation operators to build a representation of $su(2)$:
\begin{equation}
	[\hat{a}_\alpha,\hat{a}^\dagger_\beta]=\delta_{\alpha\beta}\quad\mathrm{and}\quad[\hat{a}_\alpha,\hat{a}_\beta] = [\hat{a}^\dagger_\alpha,\hat{a}^\dagger_\beta]=0\quad\mathrm{with}\quad\alpha,\beta=1,2.
\label{eq:operator-coms}
\end{equation}
The corresponding boson Fock space is identified with $\mathcal{H}_C$. The coordinate operators are realised as
\begin{equation}
	\hat{X}_i = \theta\hat{a}^\dagger_\alpha\sigma^{(i)}_{\alpha\beta}\hat{a}_\beta
\label{eq:nc-coordinates}
\end{equation}
where $\{\sigma^{(i)}\}$ are the Pauli spin matrices. The Casimir operator reads $\hat{X}^2=\hat{X}_i\hat{X}_i=\theta^2\hat{N} (\hat{N}+2)$ with $\hat{N}=\hat{a}_1^\dagger\hat{a}_1+\hat{a}_2^\dagger\hat{a}_2$ from which it is clear that each $su(2)$ representation occurs precisely once in $\mathcal{H}_C$. As a measure of radial distance we will use
\begin{equation}
	\hat{r} = \theta(\hat{N}+1).
\label{eq:distance-op}
\end{equation}

The quantum Hilbert space $\mathcal{H}_Q$ is now defined as the algebra of operators generated by the coordinates, i.e. the operators acting on $\mathcal{H}_C$ that commute with $\hat{X}^2$ and have a finite norm with respect to a weighted Hilbert-Schmidt inner product \cite{nc-well}: 
\begin{equation}
{\cal H}_Q=\left\{\psi=\sum_{j,m^\prime,m} c_{j,m^\prime,m}|j,m^\prime\rangle\langle jm|: {\rm tr_C}\left(\psi^\dagger\left(\hat{X}^2 + \theta^2\right)^{1/2}\psi\right)<\infty\right\}.
\label{eq:defineHQ}
\end{equation}
Here ${\rm tr_C}$ denotes the trace over configuration space.

The quantum mechanical angular momentum operators are realised on $\mathcal{H}_Q$ as
\begin{equation}
	\hat{L}_i\psih = \frac{\hbar}{2\theta}[\hat{X}_i, \psih]\quad{\rm with}\quad[\hat{L}_i,\hat{L}_j] = i\hbar\varepsilon_{ijk}\hat{L}_k.
\label{eq:nc-angular-ops}
\end{equation}
The simultaneous eigenstates of $\hat{L}^2$ and $\hat{L}_3$ are of the form
\begin{equation}
	\psih_{jm}=\sum_{(jm)}\frac{(\hat{a}^\dagger_1)^{m_1}(\hat{a}^\dagger_2)^{m_2}}{m_1!\,m_2!}\no{R(\hat{N})}\frac{(\hat{a}_1)^{n_1}(-\hat{a}_2)^{n_2}}{n_1!\,n_2!}
\label{eq:standard-eigenfunctions}
\end{equation}
with $j=0,1,2,\ldots$ and $m=-j,\ldots,+j$. The summation above is over the non-negative integers ($m_1,m_2,n_1,n_2$) which satisfy $m_1+m_2=n_1+n_2=j$ and $m_1-m_2-n_1+n_2=2m$. The state's radial dependence is determined by the function $R(n)=\sum_k c_k n^k$ through its normal ordered form
\begin{equation}
	\bar{R}(\hat{N})\equiv\,\no{R(\hat{N})}\,=\sum_{k=0}^\infty c_k\frac{\hat{N}!}{(\hat{N}-k)!}.
\label{eq:radial-def}
\end{equation}
\subsection{Free particle solutions}
In this, and the following two subsections we collect some results relevant to the non-commutative free particle and square well problems. Our treatment follows that of section 7 in \cite{nc-well}. We first introduce the non-commutative analogue of the Laplacian \cite{nc-coulomb} which acts on a $\hat{\psi}\in\mathcal{H}_Q$ as
\begin{equation}
 \hat{\Delta}_\theta\psih=-\frac{1}{\theta\hat{r}}[\hat{a}^\dagger_\alpha,[\hat{a}_\alpha,\psih]].
\label{eq:nc-laplacian}
\end{equation}
Using the form of $\psih_{jm}$ in \re{standard-eigenfunctions} the free particle Schr\"odinger equation 
\begin{equation}
	-\frac{\hbar^2}{2m_0}\hat{\Delta}_\theta\psih_{jm}=E\psih_{jm}
\end{equation}
can be reduced to a difference equation for $\bar{R}(n)$ on the non-negative integers, or equivalently, a differential equation for $R(n)$ on $n\in[0,\infty)$. The two solutions are related by the normal ordering procedure in \re{radial-def}. Here we only require $\bar{R}(n)$, which is found to be
\begin{equation}
	\bar{R}(n) \propto \frac{n!}{\Gamma(n+j+3/2)}P^{(j+1/2\,,\,j+1/2)}_{n}(1-\kappa^2/2)
\label{eq:jacobi-poly-free}
\end{equation}
where $\kappa = \theta k$, $E = \frac{\hbar^2k^2}{2m_0}$ and $P^{(\alpha,\beta)}_{n}(x)$ is a Jacobi polynomial. This is the non-commutative analogue of the usual Bessel function solution of the radial wave equation. By using identity (22.15.1) in \cite{abromowitz} this expression can be shown to reduce to the standard commutative result as $\theta\rightarrow0$. An important consequence of \re{jacobi-poly-free} is that it places an upper bound on the free particle energy $E$. This can be seen by considering the asymptotic behaviour of $\bar{R}(n)$ as $n\rightarrow\infty$. According to theorems  $(8.21.7)$ and $(8.21.8)$ in \cite{szego} the limit $\lim_{n\rightarrow\infty}\bar{R}(n)$ will diverge if the argument of the Jacobi polynomial falls outside the interval $[-1,1]$. For $\hat{\psi}_{jm}$ to be normalisable with respect to the trace inner product in \re{defineHQ} $\kappa$ therefore cannot exceed two, which translates into an energy upper bound of $E\leq\frac{2\hbar^2}{m_0\theta^2}$.

\subsection{The infinite fuzzy well}
We next consider a spherical well potential in non-commutative space. To this end, consider the operator
\begin{equation}
	\hat{Q} = \sum_{n=M+1}^{\infty}\sum_{k=0}^n \ket{k,\,n-k}\bra{k,\,n-k}
\label{eq:projection-ops}
\end{equation}
which projects onto the subspace of $\mathcal{H}_C$ spanned by Fock states with total particle number greater or equal to $M+1$. Through the relation between $\hat{N}$ and radial distance in \re{distance-op} this amounts to projecting onto configuration space states which are localised beyond a physical radius of $R=\theta(M+1)$. For a potential well with height $V>0$ and radius $R$ the Schr\"odinger equation now reads
\begin{equation}
	\frac{\hbar^2}{2m_0}\Delta_\theta\psih_{jm} + V\hat{Q}\psih_{jm} = E\psih_{jm}.
	\label{eq:hamiltonian-piecewise}
\end{equation}
Through \re{standard-eigenfunctions} this again translates into a difference equation for $\bar{R}(n)$. In this equation $\mathcal{M}=M-j$ is found to appear naturally as the effective radius of the well. The solution for $n\leq\mathcal{M}+1$ is just that of the free particle given in \re{jacobi-poly-free}, while for $n\geq\mathcal{M}$ it is a linear combination of non-commutative versions of the Bessel and Neumann functions. A careful analysis of the matching condition at $n=\mathcal{M},\mathcal{M}+1$ reveals that in the $V\rightarrow\infty$ limit the interior solution $\bar{R}(n)$ must vanish at $n=\mathcal{M}+1$. According to \re{jacobi-poly-free} the bound state energies of the infinite well are therefore determined by the zeroes of the Jacobi polynomials. Upon using the symmetry relation $P^{(\alpha,\alpha)}_n(-x)=(-1)^n P^{(\alpha,\alpha)}_n(x)$ and the freedom to shift the energies by a constant (thereby taking $\kappa^2/2-1$ to $\kappa^2/2$) the quantization condition reads
\begin{equation}
	P_{M-j+1}^{(j+\frac{1}{2},j+\frac{1}{2})}(\kappa^2/2) = P_{M-j+1}^{(j+\frac{1}{2},j+\frac{1}{2})}(m_0\theta^2 E /\hbar^2)  = 0.
\label{eq:jacobi-pol-inf-well}
\end{equation}
Note that, due to shifting the energies, the bound states are now arranged symmetrically around $E=0$.

\subsection{Properties of the bound state spectrum}
It follows from (\ref{eq:jacobi-pol-inf-well}) that for a specific angular momentum $j$ there are exactly $(2j+1) (M-j+1)$ bound states. This implies not only a cut-off in energy, but also in angular momentum at $j_{max}=M$. By summing over $j$ the total number of single particle states, and therefore the maximum number of spinless fermions the well can accommodate, is found to be
\begin{equation}
	\nmax=\frac{1}{6} (M+1) (M+2) (2 M+3)=\frac{M^3}{3}+\mathcal{O}(M^2).
	\label{eq:total-particles}
\end{equation}
As a consistency check on (\ref{eq:jacobi-pol-inf-well}), we derive this result in an alternative way.  Since the infinite well of radius $M$ confines particles within this radius,  the configuration space is effectively truncated at states with a  radius less or equal to $M$ and is thus finite dimensional. The states spanning this finite dimensional subspace of Fock space are $|n_1,n_2\rangle$ where $n_1+n_2=N=0,1,\ldots M$. Since the quantum Hilbert space is spanned by states of the form $|N,m^\prime,m)=|j=N/2,m^\prime\rangle\langle j=N/2,m|$ where $N=0,1,\ldots M$ and $m,m^\prime=-N/2,\ldots N/2$, the quantum Hilbert space is also finite dimensional with dimension $\sum_{N=0}^M (N+1)^2$, which agrees precisely with (\ref{eq:total-particles}). Thus, counting the number of independent solutions of the non-commutative Schr{\"o}dinger equations matches precisely the dimensionality of quantum Hilbert space.

\begin{figure}[t]
    \centering
    \begin{tabular}{c c}
    (a)&(b)\\
    \includegraphics[width=0.38\textwidth]{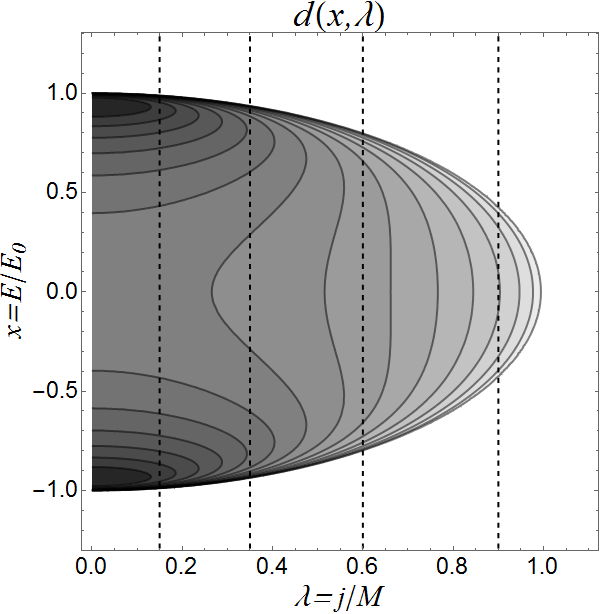}$\quad$&\includegraphics[width=0.4\textwidth]{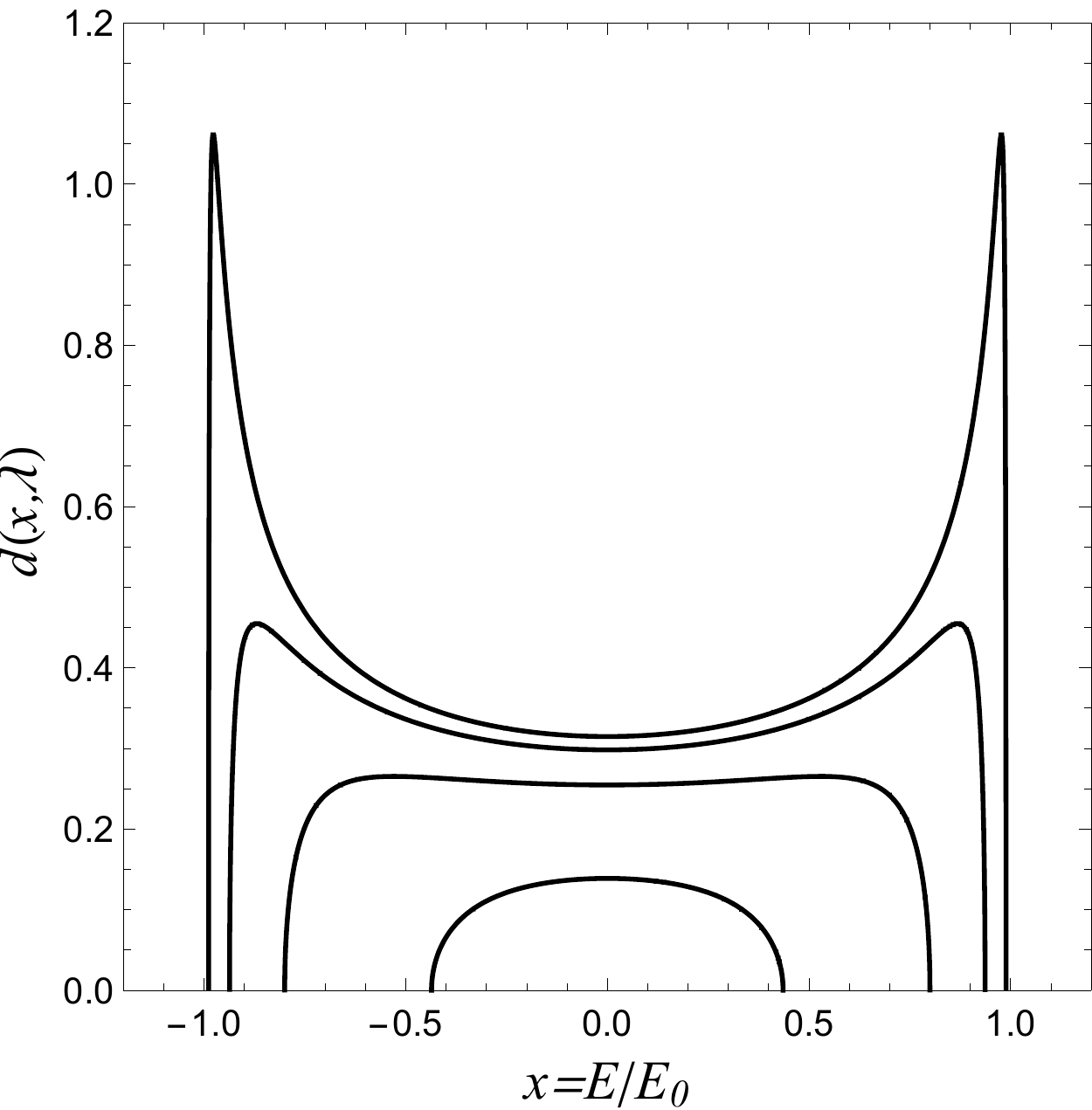}
	\end{tabular}
  \caption{Figure (a) shows a gradient plot of $d(x,\lambda)$ which defines the density of states $\rho(E,M,j)$. Figure (b) shows $d(x,\lambda)$ as a function of $x$ for the four values of $\lambda$ indicated by dashed lines in figure (a).   }
  \label{dosplots}
\end{figure}

In order to study the thermodynamics we require the density of states, which is related to the density of zeros of the Jacobi polynomial $P_{M-j+1}^{(j+1/2,j+1/2)}(x)$. The latter, in the limit of large $j$ and $M$, is $M d(x,\lambda)$ where $\lambda=j/M$ and \cite{zeros1,zeros2}
\begin{equation}
	d(x,\lambda) = \frac{\sqrt{1-x^2-\lambda^2}}{\pi\left(1-x^2\right)}.
\label{eq:dens-states}
\end{equation}
By definition $d(x,\lambda)=0$ when $|x|>\sqrt{1+\lambda^2}$. The thermodynamic density of states, still for a fixed $j$, is then $\rho(E,M,j)=(2j+1)Md(E/E_0,j/M)/E_0$ with $E_0=\frac{\hbar^2}{m_0\theta^2}$ and where $\frac{E}{E_0}$ ranges from $-\sqrt{1-\lambda^2}$ to $+\sqrt{1-\lambda^2}$. Note that the density of states is symmetric under $E\rightarrow-E$. This simple fact will have important implications for the discussion of the high/low density duality in section \ref{duality}. Also of interest is the total density of states $\rho(E,M)$ obtained by summing $\rho(E,M,j)$ over $j$. This is found to be $\rho(E,M)=M^3 d(E/E_0)/E_0$ where
\begin{equation}
	d(x) = \int_0^{\sqrt{1-x^2}}d\lambda\, 2\lambda\, d(x,\lambda)=\frac{2\sqrt{1-x^2}}{3\pi}.
\label{eq:tot-dens-states}
\end{equation}
Figure \ref{dosplots} illustrates the relevant properties of $d(x,\lambda)$ which are imprinted on the density of states $\rho(E,M,j)$. The cutoffs in the angular momentum at $j=M$ and energy at $E=E_0\sqrt{1-(j/M)^2}$ are clearly visible, as is the symmetry of the density of states around zero energy.\\

\section{The non-commutative three dimensional Fermi gas}
\label{Section 2}
\subsection{The thermodynamic $q$-potential}
\label{sec:thermo}
We now turn to the thermodynamics of a non-interacting gas of spinless fermions which are confined to an infinite spherical well with radius $R=\theta(M+1)$ in non-commutative space. We will treat this system in the grand canonical ensemble and fix the average value of the $z$-component of the total angular momentum. The grand canonical potential reads
\begin{equation}
	q(M,\tilde\beta,\tilde\mu,\tilde\omega^\prime) = \sum_{j=0}^{M}\sum_{m=-j}^{+j}\sum_{n}\log\left[1+e^{-\tilde{\beta}(-x_{n,j}-\tilde{\mu}- m\tilde{\omega}^\prime)}\right].
\label{eq:3D-q-pot-discrete2}
\end{equation}
Here $\{x_{n,j}\}$ is the set of $M-j+1$ zeros of the Jacobi polynomial $P_{M-j+1}^{(j+1/2,j+1/2)}(x)$, as is required by the quantisation condition in \re{jacobi-pol-inf-well}. We have also introduced the non-commutative energy scale $E_0\equiv\frac{\hbar^2}{m_0 \theta^2}$ in terms of which we defined the dimensionless inverse temperature $\tilde{\beta}=E_0\,\beta$ and chemical potential $\tilde\mu=\frac{\mu}{E_0}$.  The dimensionless angular velocity $\tilde\omega^\prime\equiv \frac{\hbar\omega}{E_0}$ serves as a Lagrange multiplier for fixing the dimensionless total $z$-component of the angular momentum $\tilde L_3^{\rm tot}\equiv\frac{L_3^{\rm tot}}{\hbar}$. Throughout we follow the convention that quantities denoted by a tilde are dimensionless.

In the thermodynamic (large $M$) limit the sums over $j$ and $m$ may be replaced by integrals over $\lambda = j/M$ and $\alpha = m/M$ while the $n$ summation becomes $\int dx\,d(x,\lambda)$ with $d(x,\lambda)$ given in \re{dens-states}. This produces
\begin{equation}
 q(M,\tilde\beta,\tilde\mu,\tilde\omega) = M^3\int_{0}^{1}d\lambda\int_{-\lambda}^{+\lambda}d\alpha\int_{x_{-}(\lambda)}^{x_{+}(\lambda)}dx\,d(x,\lambda)\log\left[1+e^{-\tilde{\beta}(x-\tilde{\mu}-\alpha\tilde\omega)}\right]
\label{eq:3D-q-pot-cont}
\end{equation}
where $\tilde\omega = M \tilde\omega^\prime$ and $x_\pm(\lambda)=\pm\sqrt{1-\lambda^2}$. This expression serves as the starting point for calculating the various thermodynamic quantities. These are obtained from $q$ using
\begin{equation}
 N =\frac{1}{\tilde\beta} \left(\pder{q}{\tilde\mu}\right)_{\tilde T,\tilde\omega,\tilde V} \quad\quad\quad \ltot=\frac{M}{\tilde\beta}\left(\pder{q}{\tilde\omega}\right)_{\tilde T,\tilde \mu,\tilde V} \quad\quad\quad \frac{S}{k} = q-\tilde\beta\left(\pder{q}{\tilde\beta}\right)_{\tilde\omega,\tilde\mu,\tilde V} \quad\quad\quad \tilde{P} = \frac{q}{\tilde\beta \tilde V}.
\end{equation}
Here $S$ is the entropy, $N$ the number of particles, $\tilde{P}=\frac{4\pi\theta^3}{E_0}P$ a dimensionless pressure and $\tilde V=\frac{M^3}{3}\approx\frac{V}{4\pi\theta^3}$ the dimensionless system volume.  The latter is also (to leading order in $M$) the total number of single particle states available to the system and thus represents the maximum number of spinless particles that can be accommodated by the system.  It is therefore convenient to define 
\begin{equation}
\label{maxnum}
N_{\rm max}\equiv \tilde V=\frac{M^3}{3}\approx\frac{V}{4\pi\theta^3}.  
\end{equation}
Note that the physical meaning of $N_{\rm max}$ is the total number of cells with volume $V_0\equiv 4\pi\theta^3$ that fits into the volume $V$.   It is therefore also useful and sensible to define a maximum density, which is the density obtained when each cell is occupied by exactly one particle, i.e.
\begin{equation}
\rho_{\rm max}\equiv\frac{1}{V_0}=\frac{1}{4\pi\theta^3}.
\end{equation}
In terms of $\rho_{\rm max}$ the dimensionful density reads
\begin{equation}
\rho=\frac{N}{V}=\rho_{\rm max}\tilde\rho
\end{equation}
where $\tilde\rho=\frac{N}{\tilde V}$ is the dimensionless density.

We introduce one further notion that will turn out to be useful later.  Suppose we fix the number of particles in the system to be $N$.  Then there is  a minimum volume that can accommodate this number of particles given by 
\begin{equation}
\label{minvol}
N\approx \tilde{V}_{\rm min}=\frac{M_{\rm min}^3}{3}=\frac{V_{\rm min}}{4\pi\theta^3}.
\end{equation}
Here $M_{\rm min}$ is the smallest integer larger than or equal to $(3N)^{1/3}$, so that the equality between $N$ and $\tilde{V}_{\rm min}$ holds to $O(1/M_{\rm min})$.  In further applications where $N$ and $M_{\rm min}$ are assumed to be large, this correction will be ignored.

It will often be convenient to work in terms of the fugacity $z=e^{\tilde\beta\tilde\mu}=e^{\beta\mu}$.  In terms of this the $q$-potential of (\ref{eq:3D-q-pot-cont}) reads
\begin{equation}
	 q(M,\tilde\beta,z,\tilde\omega) = M^3\int_{0}^{1}d\lambda\int_{-\lambda}^{+\lambda}d\alpha\int_{x_{-}(\lambda)}^{x_{+}(\lambda)}dx\,d(x,\lambda)\log\left[1+ze^{-\tilde{\beta}(x-\tilde\omega\alpha)}\right],
\label{eq:3D-q-pot-cont-fug}
\end{equation}
where $z<<1$ corresponds to the low density limit and $z>>1$ the high density limit. The corresponding thermodynamic identities are then given by 
\begin{equation}
 N =z \left(\pder{q}{z}\right)_{\tilde T,\tilde\omega,\tilde V} \quad\quad\quad \ltot=\frac{M}{\tilde\beta}\left(\pder{q}{\tilde\omega}\right)_{\tilde T,z,\tilde V} \quad\quad\quad \frac{S}{k} = q-N\log z-\tilde\beta\left(\pder{q}{\tilde\beta}\right)_{\tilde\omega,z,\tilde V} \quad\quad\quad \tilde{P} = \frac{q}{\tilde\beta \tilde V}.
\label{eq:3D-thermo-quant}
\end{equation}

\subsection{The high/low density duality}
\label{duality}

Using the symmetry of the density of states under $x\rightarrow -x$ one can trivially rewrite (\ref{eq:3D-q-pot-cont-fug}) as
\begin{equation}
q(M,\tilde\beta,z,\tilde\omega) = \frac{M^3}{3}\log z+q(M,\tilde\beta,z^{-1},\tilde\omega).
\label{eq:3D-q-pot-cont-dual}
\end{equation}
This demonstrates a remarkable duality between the high and low density $q$-potentials, i.e., we can express the high density $q$-potential ($z>>1$) in terms of the low density $q$-potential ($z<<1$).  This duality, in turn, is a direct consequence of the infra-red/ultraviolet duality of the density of states.  It is therefore sufficient to compute the $q$-potential in the low density limit $z<<1$ and infer the high density behaviour from the above duality.

\section{Thermodynamics of the non-commutative three dimensional Fermi gas}
\label{Section 3}

In this section we compute the thermodynamics of the non-commutative Fermi gas. Given the high/low density duality above, our strategy will be to first compute the $q$-potential in the low density limit and then derive the high density behaviour.  It is also convenient to consider the unconstrained ensemble with $\tilde\omega=0$, and therefore $\ltot=0$, and the constrained ensemble with $\ltot\ne 0$ separately.

\subsection{The unconstrained ensemble with $\tilde\omega=0$ and $L_3^{\rm tot}=0$.}
We start by considering the unconstrained case with $\tilde\omega=0$, which corresponds to an average angular momentum $L_3^{\rm tot}=0$. In this case the $q$-potential reads
\begin{equation}
	 q(M,\tilde\beta,z) = \frac{2M^3}{3\pi}\int_{-1}^1dx\,\sqrt{1-x^2}\log\left[1+ze^{-\tilde{\beta}x}\right],
\label{eq:3D-q-pot-cont-uncon}
\end{equation}
which follows from combining \eqref{eq:dens-states} and \eqref{eq:tot-dens-states} in \eqref{eq:3D-q-pot-cont}.

\label{sec:l0-system}
\subsubsection{The low density limit.}

In this limit $z<<1$, but we still have to consider two cases.  In the first case the dimensionless chemical potential $\tilde\mu=\frac{1}{\tilde\beta}\log{z}<-1$ and in the second case $0>\tilde\mu=\frac{1}{\tilde\beta}\log z>-1$ where, for a low density, $\tilde\mu\approx -1$.

We begin by considering $\tilde\mu=\frac{1}{\tilde\beta}\log{z}<-1$ in the high and low temperature limits, starting with the high temperature limit when $\tilde\beta<<1$.   After solving the fugacity to second order in the density, which is required to detect the lowest order non-commutative corrections to the pressure, we obtain for the central thermodynamic quantities to the same order in the density, lowest order in $\tilde{\beta}$  and after restoring dimensions
\begin{eqnarray}
 && z=\frac{\rho}{\rho_{\rm max}}\left(1+\frac{\rho}{\rho_{\rm max}}\right)<<1,\nonumber\\
 &&P=\rho kT\left(1+\frac{\rho}{2\rho_{\rm max}}\right),\nonumber\\
 &&S=Nk\left[1-\log\left(\frac{\rho}{\rho_{\rm max}}\right)-\frac{\rho}{2\rho_{\rm max}}\right],
\label{hightld}
\end{eqnarray}
where $\rho$ is the dimensionful density.  Note that the dimensionful temperature here corresponds to $kT>>E_0$ so that these are ultra high temperatures and not physically realistic.  It is therefore also not surprising that non-commutative effects feature explicitly as the thermodynamics will be sensitive to the high energy cut-off $E_0$ at these high temperatures.

Let us now consider the more physical limit of low temperature when $\tilde\beta>>1$.   Following the same procedure as above, we find for the central thermodynamic quantities up to second order in density, first order in $kT$ and after restoring dimensions
\begin{eqnarray}
&&z=e^{-\frac{E_0}{kT} }\lambda ^3 \rho\left[ \left(1+\frac{3 kT}{8E_0} \right)+ \frac{\lambda ^3 \rho }{2 \sqrt{2}}\left(1+\frac{15kT }{16E_0 }\right)\right],\nonumber\\
&& P=\rho kT\left[1+\frac{\sqrt{2}\lambda^3\rho}{8}\left(1+\frac{9kT}{16E_0}\right)\right],\nonumber\\
&& S=Nk\left[\frac{5}{2}-\log\left(\lambda^3\rho\right)+\frac{\lambda^3\rho}{8\sqrt{2}}\left(1-\frac{9kT}{16E_0}\right)\right].
\label{hightld1}
\end{eqnarray}
Here $\lambda=\frac{h}{\sqrt{2\pi m_0 kT}}$ is the standard thermal length.  Note that the higher order corrections in the density consists of ordinary quantum corrections that are also present in the commutative gas as well as non-commutative corrections that exhibit an explicit dependence on the non-commutative energy scale $E_0$.  Also note that the dimensionless chemical potential is given by 
\begin{equation}
\tilde\mu=\frac{1}{\tilde\beta}\log{z}=-1+\frac{1}{\tilde\beta}\log\left(\lambda ^3 \rho\right)+\log\left[\left(1+\frac{3}{8\tilde\beta} \right)+ \frac{\lambda ^3 \rho }{2 \sqrt{2}}\left(1+\frac{15 }{16\tilde\beta }\right)\right].
\end{equation}
The above approximation is only valid when $\tilde\mu<-1$, which places constraints on the value of $\lambda^3\rho$ for the validity of the approximation.  In particular note that one cannot take the zero temperature limit before the zero density limit.

Finally we consider the case when $0>\tilde\mu=\frac{1}{\tilde\beta}\log z>-1$ and low temperature $\tilde\beta>>1$.  The way to proceed in this case is to perform a Sommerfeld expansion, which we implement here in the following form \cite{pathria}:
\begin{equation}
\int_a^b dx f(x)\log(1+e^{-\beta (x-\mu)})=\beta\int_a^\mu dxf(x)(\mu-x)+\frac{\pi^2}{6\beta}f(\mu)+\ldots
\end{equation}
This yields 
\begin{equation}
\label{qlowlow1}
q=\frac{2 M^3}{3 \pi } \left(\frac{\pi  \sqrt{1-\tilde\mu ^2}}{6 \tilde\beta }+\frac{ \tilde\beta}{12}  \left(2 \sqrt{1-\tilde\mu ^2} \left(\tilde\mu ^2+2\right)+3 \pi  \tilde\mu +6 \tilde\mu  \sin ^{-1}(\tilde\mu )\right)\right).
  \end{equation}
To solve for the chemical potential we write $\tilde\mu$ as $\tilde\mu=-1+\tilde\mu_0+\frac{\tilde\mu_2}{\tilde\beta^2}$ and expand the $q$-potential in orders of $\tilde\beta^{-1}$ and $\tilde\mu_0$ up to second order, which assumes a low density.  This yields
\begin{equation}
\tilde\mu=-1+\tilde{\epsilon}_F-\frac{\pi^2}{12\tilde{\beta}^2\tilde{\epsilon}_F}\left(1-\frac{\tilde\epsilon_F}{2}\right),
\label{chemp}
\end{equation}
where $\epsilon_F=E_0\tilde\epsilon_F=(\frac{3N}{4\pi V})^{2/3}\frac{h^2}{2m_0}$ is the Fermi energy.  This expansion is only valid if $\tilde\epsilon^2_F>\frac{\pi^2}{12\tilde\beta^2}\left(1-\frac{\tilde\epsilon_F}{2}\right)$. The thermodynamics now follows as usual and after restoring dimensions we find
\begin{eqnarray}
&&P=\frac{2}{5}\rho\epsilon_F\left[1+\pi^2(kT)^2\left(\frac{5}{12\epsilon_F^2}-\frac{1}{48E_0\epsilon_F}-\frac{29}{896 E_0^2}\right)\right], \nonumber\\
&&S=\frac{kN\pi^2}{2\beta\epsilon_F}\left(1-\frac{\epsilon_F}{4E_0}-\frac{\epsilon_F^2}{32E_0^2}\right).
\label{therm}
\end{eqnarray}

Finally we remark that all the results above agree with the results for a commutative perfect Fermi gas \cite{pathria} when $\theta\rightarrow 0$ and $E_0\rightarrow\infty$.

\subsubsection{The high density limit.}

Now that we have established the low density thermodynamics, we can construct the thermodynamic quantities at high density straightforwardly through the use of the low/high density duality.  In particular we note from (\ref{eq:3D-q-pot-cont-dual}) 
\begin{equation}
z^{-1}\frac{\partial q(M,\tilde\beta,z^{-1},\tilde\omega)}{\partial z^{-1}}=\frac{M^3}{3}-z\frac{\partial q(M,\tilde\beta,z,\tilde\omega)}{\partial z}=\frac{M^3}{3}-N\equiv N_h.
\label{holefug}
\end{equation}
Here $N_h$ is the number of holes or unfilled states.  This implies that the thermodynamics at high density is determined by the thermodynamics of a gas of holes at low density and in particular the fugacity of the low density gas of holes is related to the inverse fugacity of the dense gas of particles, i.e.,  $z_h\equiv z^{-1}$. The essential difference between the high and low density limits resides in the logarithmic term of the fugacity where the inverse fugacity is determined by $N_h$ through (\ref{holefug}).  Since we have already computed the particle fugacity at low density, we can simply read off the corresponding inverse or hole fugacity at low density by replacing $N$ by $N_h$ and correspondingly $\tilde\rho$, $\rho$ by $\tilde\rho_h=\frac{N_h}{\tilde V}$ and $\rho_h=\frac{N_h}{V}$, respectively.  Also note the following relations between the dimensionless and dimensionful particle and hole densities
\begin{eqnarray}
&&\tilde\rho+\tilde\rho_h=1,\nonumber\\
&& \rho+\rho_h=\rho_{\rm max}.
\label{rel}
\end{eqnarray}
The dimensionful pressure and entropy of the high density gas can now be expressed quite simply in terms of the corresponding quantities of the low density hole gas.  In particular we have from (\ref{eq:3D-q-pot-cont-dual}) and the thermodynamic relations (\ref{eq:3D-thermo-quant})
\begin{eqnarray}
&& P(M,\tilde\beta,z,\tilde\omega)=-\rho_{\rm max}kT \log{z_h}+ P(M,\tilde\beta,z_h,\tilde\omega),\nonumber\\
&& S(M,\tilde\beta,z,\tilde\omega)= S(M,\tilde\beta,z_h,\tilde\omega).
\label{highpent}
\end{eqnarray}

From these results it is simple to write down the high density thermodynamics dual to the low density cases studied above.  One simply applies the results of the low density case with the replacements $N\rightarrow N_h$ and $\rho\rightarrow\rho_h$.  The only additional term, which contains the essential physics at high density, appears in the pressure as a logarithm. 

Let us consider the high temperature and high density case where $\tilde\mu=\frac{1}{\tilde\beta}\log{z}>1$.  As the duality $z\leftrightarrow z^{-1}$ translates on the level of the chemical potential to $\tilde\mu\leftrightarrow-\tilde\mu$, this implies for the dual low density gas of holes $\tilde\mu=\frac{1}{\tilde\beta}\log{z}<-1$.  We can then use the results of  (\ref{hightld}) and (\ref{highpent}) to write down the hole fugacity, pressure and entropy explicitly.  However, the detailed expressions are not relevant for our current discussion and we focus here only on the dominant terms that highlight the effects of non-commutativity.  Other terms, corresponding to the contribution from the low density gas of holes and correction terms that arise from expanding the logarithm to second order in the hole density, are represented by dots in the expressions that follow.  We find
\begin{equation}
 P(M,\tilde\beta,\rho,\tilde\omega)=-\rho_{\rm max}kT\log\left(1-\frac{\rho}{\rho_{\rm max}}\right)+\ldots
\end{equation}
Here the leading logarithmic term has been written explicitly in terms of the particle density using (\ref{rel}).  We note that the pressure diverges logarithmically at the maximum density.  It is also insightful to make the volume dependence of the pressure explicit at fixed particle number by noting that $\frac{\rho}{\rho_{\rm max}}=\frac{V_{\rm min}}{V}$:
\begin{equation}
 P(M,\tilde\beta,V,\tilde\omega)=-\rho_{\rm max}kT\log\left(1-\frac{V_{\rm min}}{V}\right)+\ldots
\end{equation}
Again the divergence in the pressure is evident as the minimum volume is approached.  This implies that these systems cannot be compressed beyond a minimum size, which corresponds to the minimum number of states required to accommodate the number of particles.

Similarly we find in the case of high density $\tilde\mu=\frac{1}{\tilde\beta}\log{z}>1$ and low temperature from (\ref{hightld1}) and (\ref{highpent})
\begin{equation}
 P(M,\tilde\beta,\rho,\tilde\omega)=\rho_{\rm max}E_0-\rho_{\rm max}kT\log\left(\lambda^3(\rho_{\rm max}-\rho)\right)+\ldots
\end{equation}
Again the pressure exhibits a logarithmic divergence as the density approaches the maximum density.  It is also important to note the large temperature independent degeneracy pressure that contributes even at zero temperature.  The volume dependence at fixed particle number can also be made explicit as above:
\begin{equation}
P(M,\tilde\beta,V,\tilde\omega)=\rho_{\rm max}E_0-\log\left(\lambda^3\rho_{\rm max}\right)-\rho_{\rm max}kT\log\left(1-\frac{V_{\rm min}}{V}\right)+\ldots
\end{equation}
Once again the pressure diverges at the minimum volume, exhibiting the incompressibility of the system as the minimum size is approached.

Finally we consider the low temperature and high density case when $0<\tilde\mu=\frac{1}{\tilde\beta}\log z<1$, which implies for the dual low density gas of holes $0>\tilde\mu=\frac{1}{\tilde\beta}\log z>-1$.  Recognising the logarithmic term in the duality relation as $\tilde\beta\tilde\mu=-\tilde\beta\tilde\mu_h$, the pressure and entropy can be written down immediately by collecting the results from (\ref{chemp}) and (\ref{therm}).  We only show the logarithmic contribution to the pressure explicitly
\begin{equation}
P(M,\tilde\beta,V,\tilde\omega)=\rho_{\rm max}\left(E_0-{\epsilon}^h_F+\frac{\pi^2 (kT)^2}{12{\epsilon}^h_F}\left(1-\frac{\epsilon^h_F}{2E_0}\right)\right)+\ldots
\end{equation}
Here $\epsilon^h_F=(\frac{3N_h}{4\pi V})^{2/3}\frac{h^2}{2m_0}$ is the Fermi energy of holes. We note again the presence of a temperature independent degeneracy pressure.  Some care is required to investigate the behaviour of the pressure when the density tends to the maximum density or, equivalently, when the hole density and thus the hole Fermi energy tends to zero.  One cannot take this limit before the zero temperature limit as the validity of the approximations made here is then violated.  However, one can take the zero temperature limit followed by the zero hole density limit, in which case the pressure shows no divergence.

The high/low density duality has important repercussions for the entropy at high density.  In particular we note from (\ref{highpent}),  (\ref{hightld}) and (\ref{hightld1}) that the entropy at high density is extensive in the number of holes, $N_h$, rather than the number of particles.  This implies that the entropy at high density generally exhibits a non-extensive behaviour in the particle number, mass and volume of the system.  The precise dependence of the entropy on these quantities will generally depend on the physical process under consideration.  One such process is one in which the system is compressed by an external pressure to a finite number, $n$, of radial units (recall that the radius is quantised as in (\ref{eq:distance-op})) larger than the minimum radius $M_{\rm min}$ (see (\ref{minvol})).  Particles are now added to the system leading to an increased minimum size, but the size of the system is maintained at the same number of radial units above the increasing minimum size by adjusting the external pressure. Under these conditions the number of holes is given by 
\begin{equation}
N_h=\frac{M^3}{3}-N=\tilde V-\tilde V_{\rm min}=\frac{(M_{\rm min}+n)^3}{3}-\frac{(M_{\rm min})^3}{3}=nM_{\rm min}^2+O(M_{\rm min}^{-1})=\frac{n A_{\rm min}}{4\pi\theta^2}.
\end{equation}
Here $A_{\rm min}$ is the dimensionful area of the minimal including volume.  For this process, this is therefore the quantity on which the entropy depends extensively at high density and thus the entropy will exhibit an area law rather than a volume dependence at high density. 

Here our interest is in the behaviour at low temperatures.  In addition we are interested in the extremal high density case where the hole density is low, which corresponds to a chemical potential $\tilde\mu>1$.  Then (\ref{hightld1}) applies to the gas of holes and from (\ref{highpent}) we can write the entropy to leading order in $A_{\rm min}$ 
\begin{equation}
S=\frac{5nk}{8\pi\theta^2}A_{\rm min}.
\end{equation}

In section \ref{Section 4} we consider a gas confined by gravity, which results in a different non-extensive behaviour for the entropy.

\subsection{Numerical results}

In this section we present exact numerical computations for the thermodynamic quantities described in the previous section, compare with the various approximations and point out the generic features of these thermodynamic quantities, with emphasis on the non-commutative effects.

\begin{figure}[t]
    \centering
    \begin{tabular}{c c}
    (a)&(b)\\
    \includegraphics[width=0.40\textwidth]{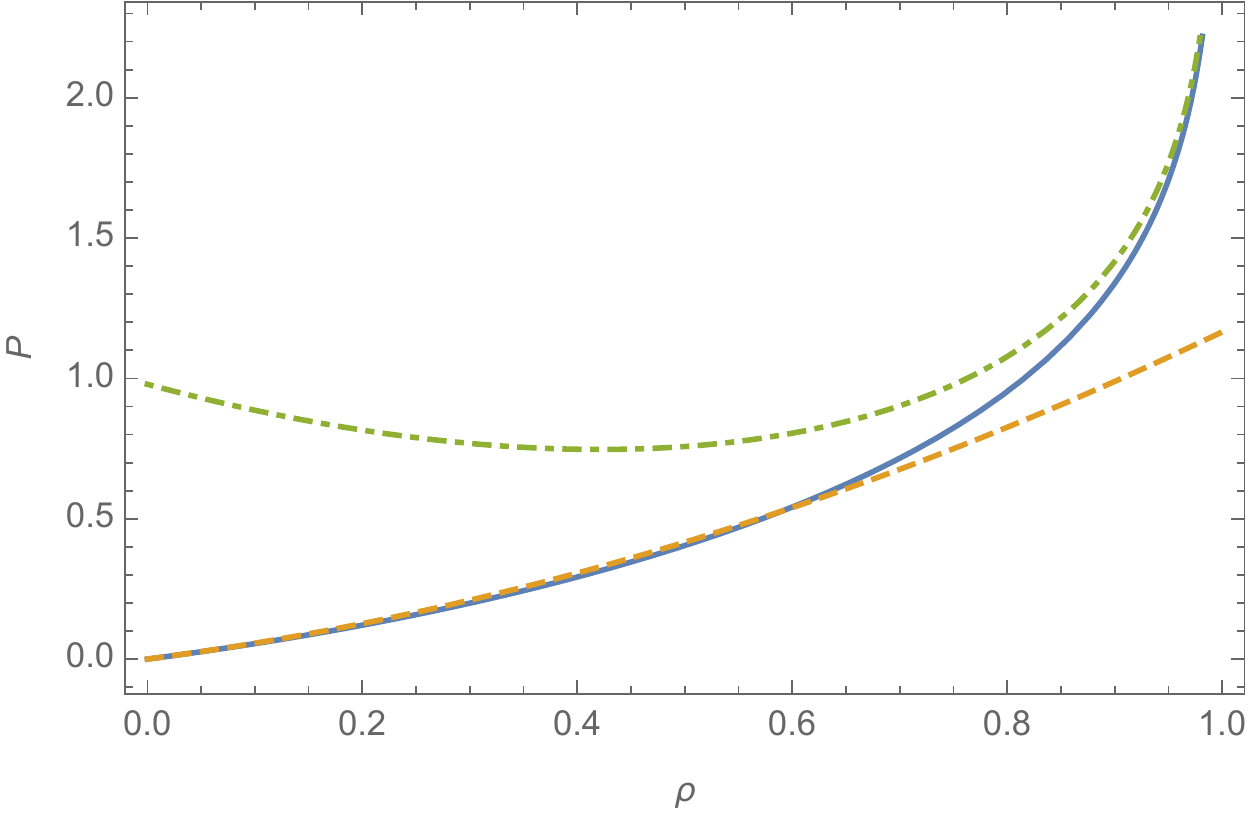}$\quad$&\includegraphics[width=0.40\textwidth]{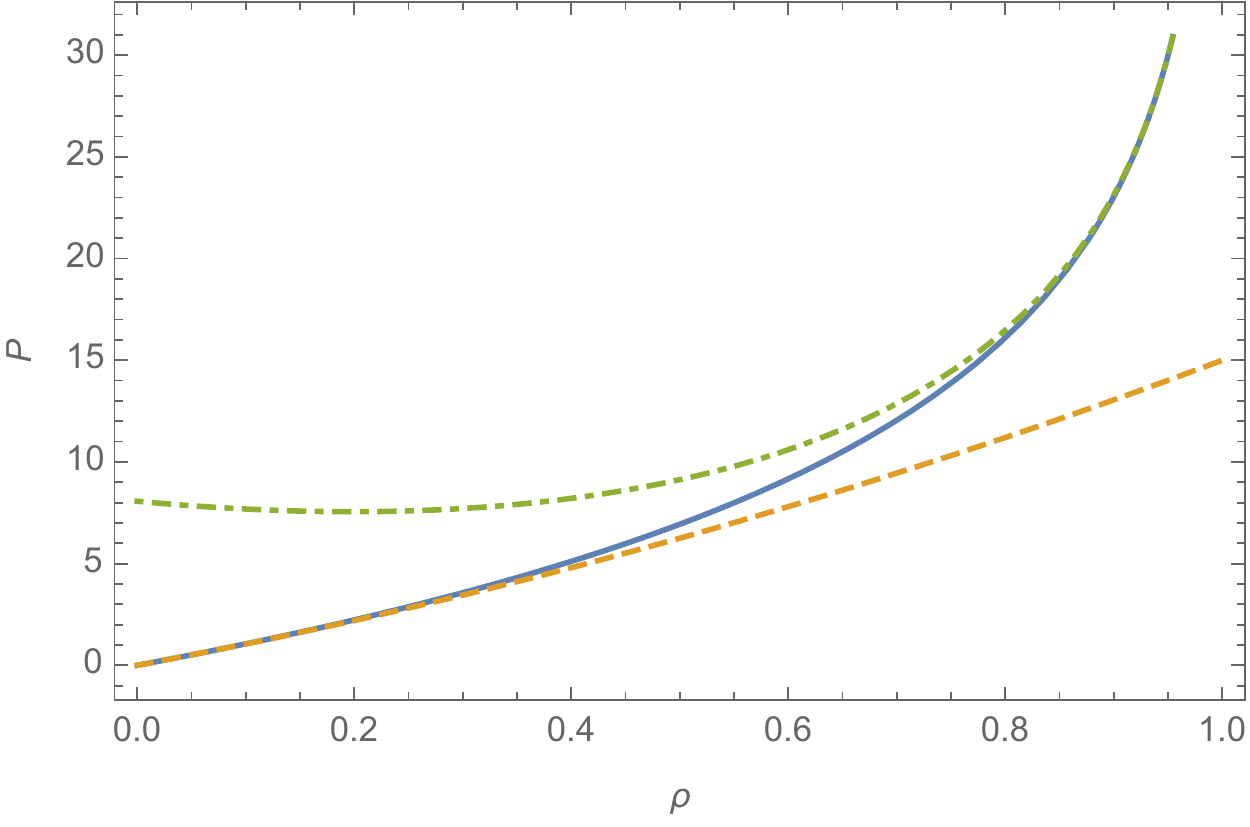}\\
	(c)&(d)\\
	\includegraphics[width=0.40\textwidth]{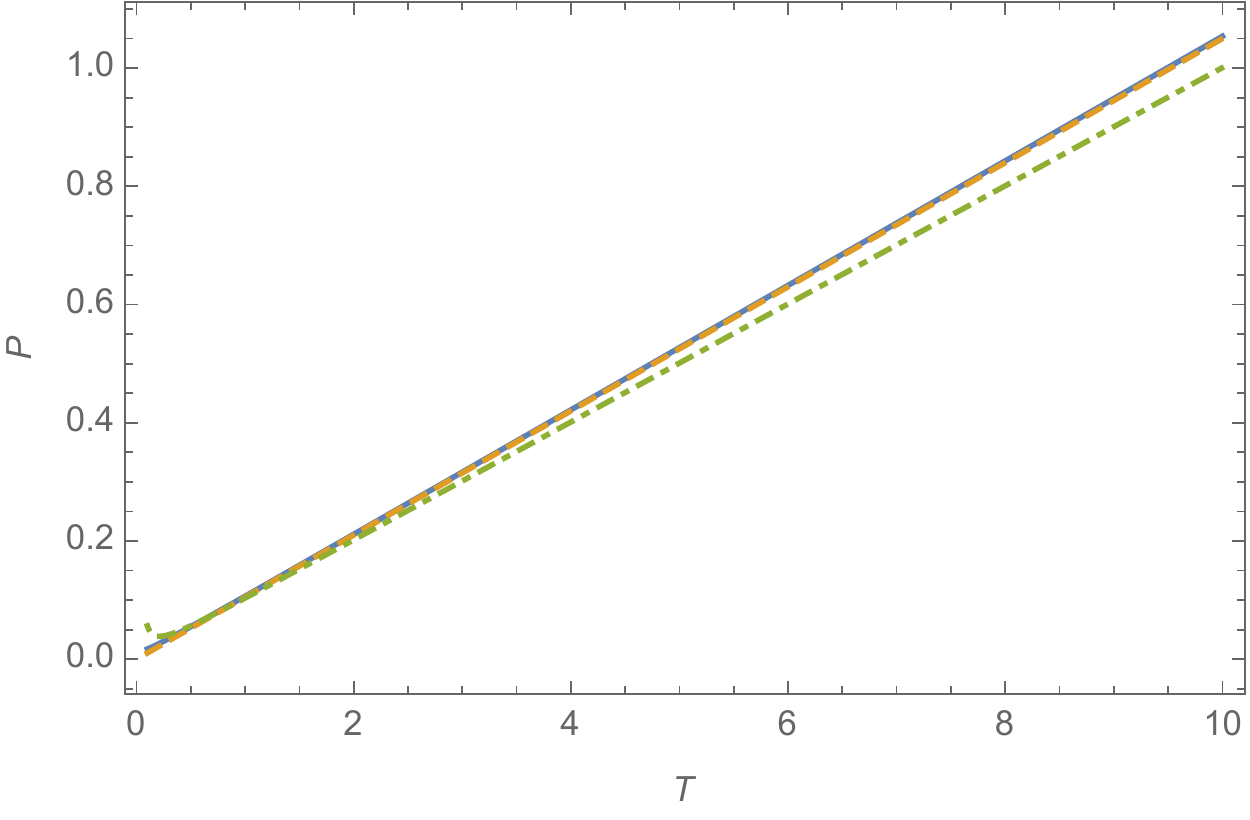}$\quad$&\includegraphics[width=0.40\textwidth]{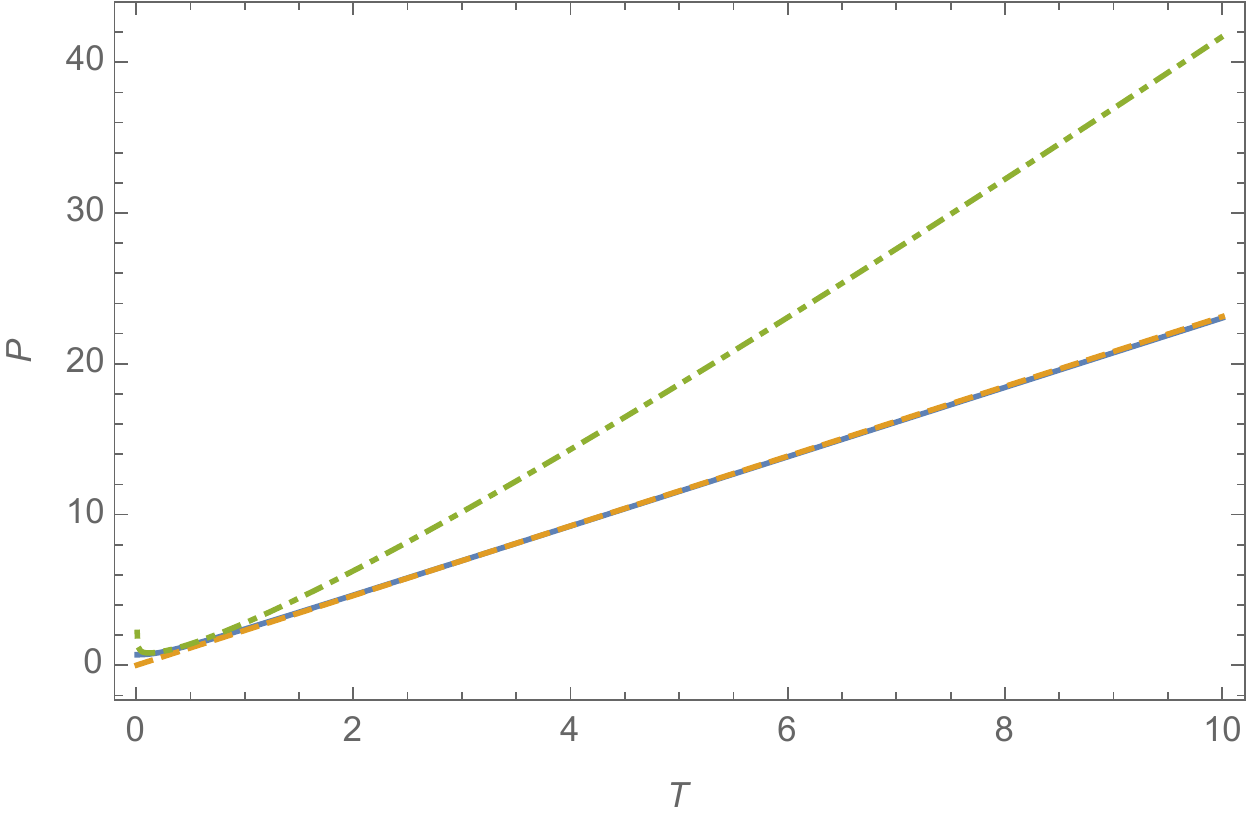}\\
	   (e)&(f)\\
    \includegraphics[width=0.40\textwidth]{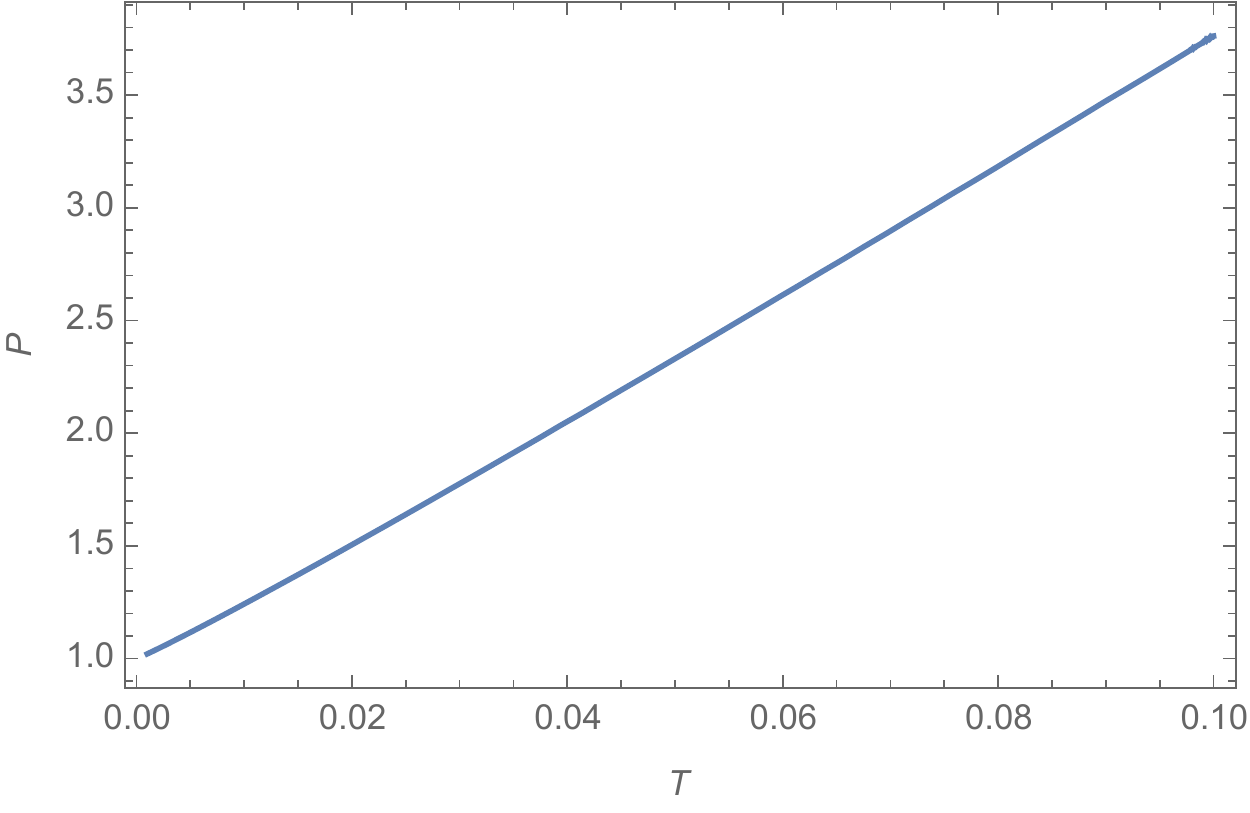}$\quad$&\includegraphics[width=0.40\textwidth]{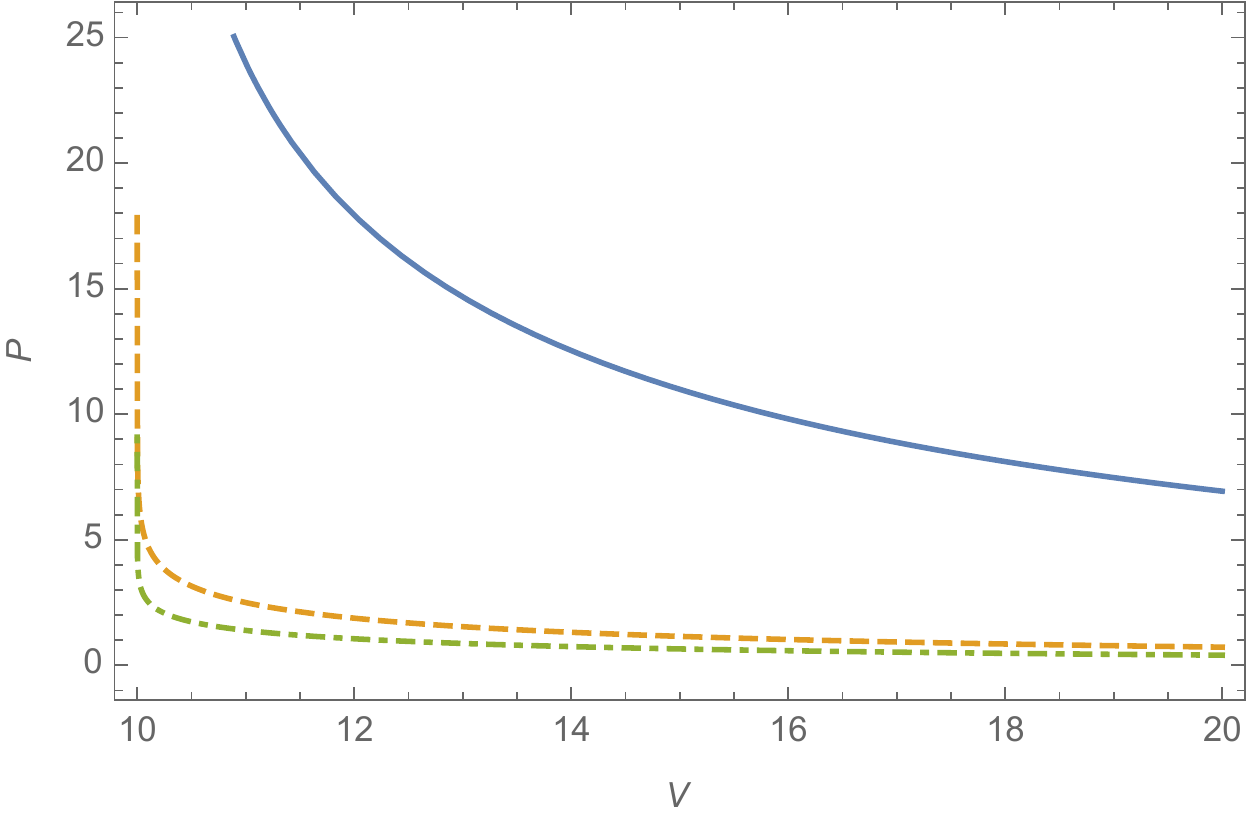}\\
\end{tabular}
  \caption{Summary of results for dimensionless pressure.  Figure (a) shows the dependence of pressure on the density (solid line) at low temperature $\tilde\beta=2$.  Also shown are the low density (dashed line) and its dual high density (dash-dotted line) approximations at low temperature. (b) Shows the same result (solid line) at high temperature $\tilde\beta=0.1$.  (c) Shows the temperature dependence of the pressure (solid line) at low density.  Also shown are the high (dashed line) and low temperature,  low density (dash-dotted line) approximations.  (d) Shows the same result at high density.  (e) Shows the exactly computed pressure  at high density ($\tilde\rho=0.99$).  One notes that the pressure saturates at the degenerate value of $\tilde P=1$ at low temperature.  (f) Shows the volume dependence of the pressure at fixed temperatures ($\tilde\beta=0.1$ (solid line), $\tilde\beta=1$ (dashed line) and $\tilde\beta=2$ (dash-dotted line)) for a minimum system size $\tilde V_{\rm min}=10$.}
  \label{figpressure}
\end{figure}

\begin{figure}[t]
    \centering
    \begin{tabular}{c c}
    (a)&(b)\\
    \includegraphics[width=0.40\textwidth]{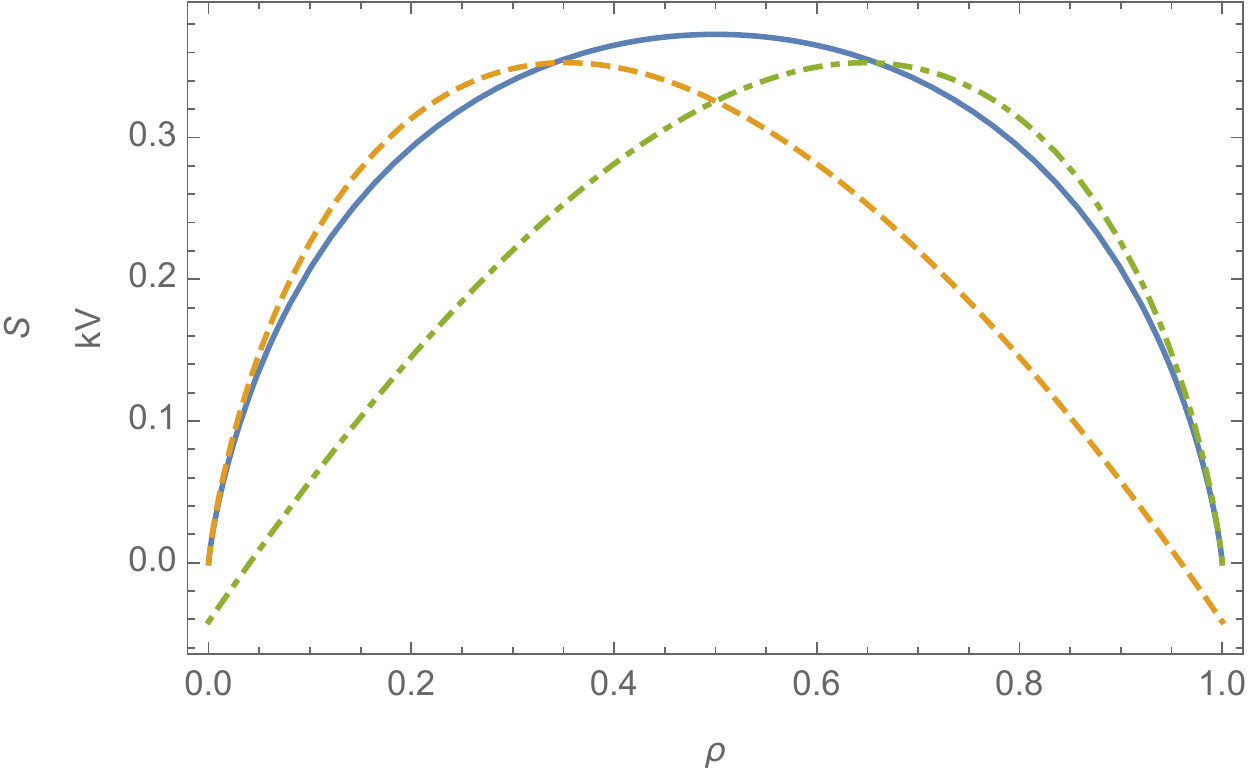}$\quad$&\includegraphics[width=0.40\textwidth]{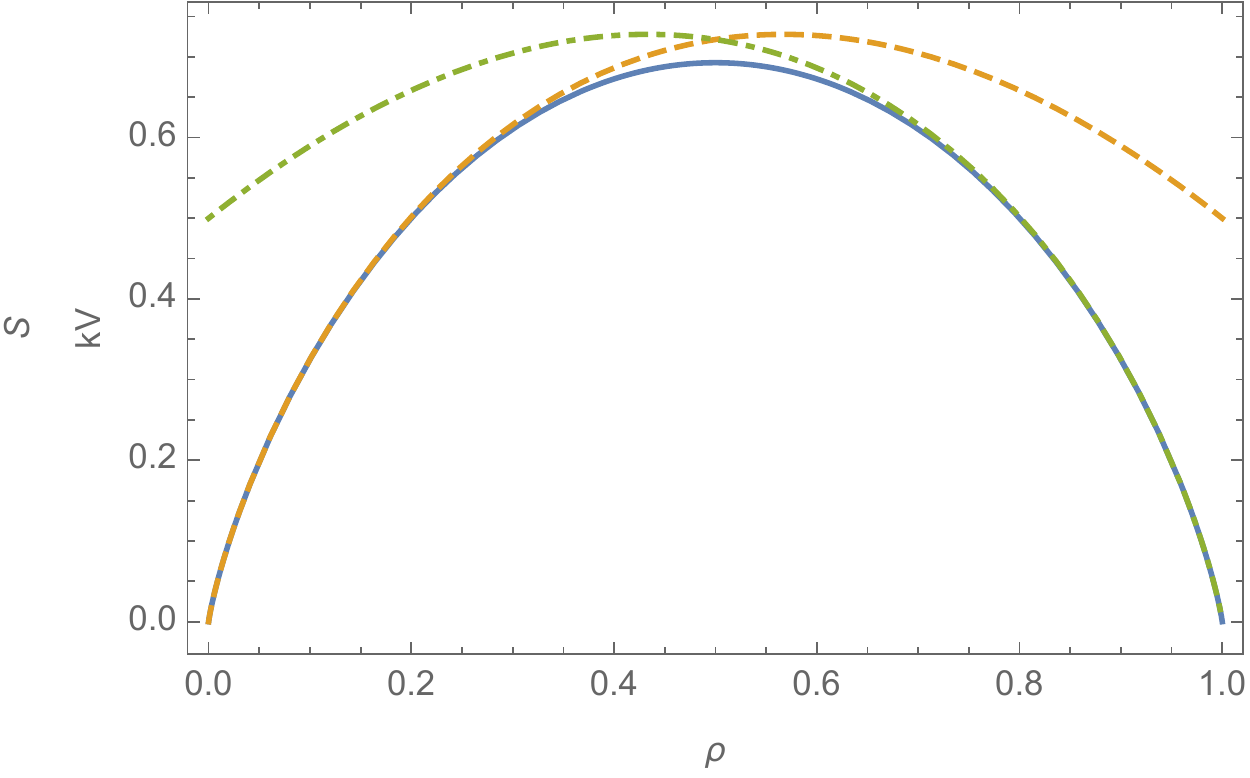}\\
      (c)&(d)\\
    \includegraphics[width=0.40\textwidth]{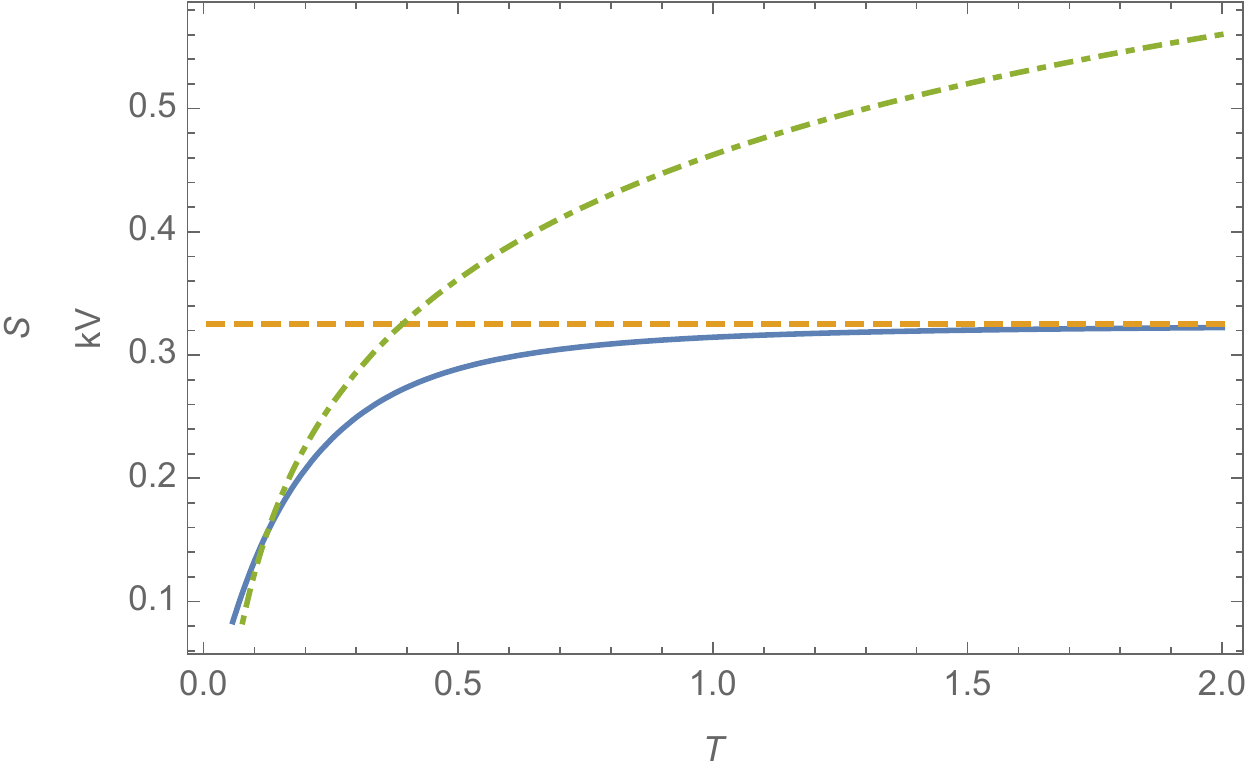}$\quad$&\includegraphics[width=0.40\textwidth]{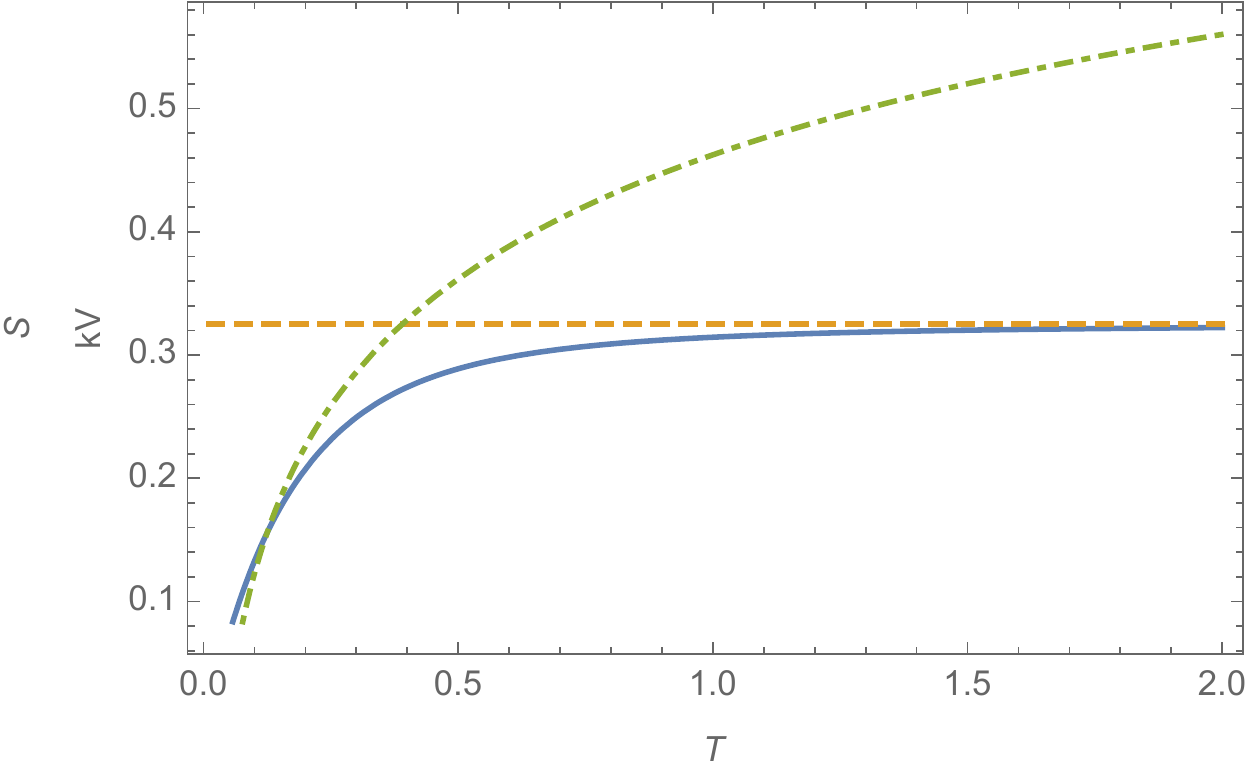}\\
\end{tabular}
  \caption{Summary of results for dimensionless entropy.  The quantity shown is $\frac{S}{k\tilde V}$.  Figure (a) shows this at low temperature ($\tilde\beta=5$).  Also shown are the low (dashed line) and high (dash-dotted line) density approximations at low temperature. (b) Corresponding result at high temperature ($\tilde\beta=0.1$).  Also shown are the low (dashed line) and high (dash-dotted line) density approximations at high temperature. Figures (c) and (d) show the corresponding results at low ($\tilde\rho=0.1$) and high densities ($\tilde\rho=0.9$) as a function of the dimensionless temperature $T=\frac{1}{\tilde\beta}$.  In figure (c) the high (dashed) and low (dash-dotted) temperature approximations at low density and in (d) the high (dashed) and low (dash-dotted) temperature approximations at high density are also shown.}
    \label{entropy}
\end{figure}

Figures (\ref{figpressure}) and (\ref{entropy}) summarise our results. All quantities shown in these graphs are dimensionless. Figures (\ref{figpressure}a) and (\ref{figpressure}b) show the dependence of the pressure on the density at low and high temperatures, respectively.  Shown on these graphs are also the approximations based on equations (\ref{hightld}) and (\ref{hightld1}) and their high density duals.  Overall there is good agreement between the exact and approximate results in their region of validity.  Figures (\ref{figpressure}c) and (\ref{figpressure}d) display the same information, but now for the temperature dependence of the pressure.  Figure (\ref{figpressure}e) demonstrates the saturation of pressure at high density and low temperature at the temperature independent dimensionless degeneracy pressure $\tilde P=1$.  Figure (\ref{figpressure}f) shows the dependence of the pressure on volume for three different temperature values.  Here a dimensionless minimum volume of $\tilde V_{\rm min}=10$ was chosen.  In all cases the incompressibility at minimum system size is apparent.  Also note that these curves represent the isotherms.  

Figures (\ref{entropy}a) and (\ref{entropy}b) show the dependence of the entropy on the density at fixed volume $\tilde V$.  The quantity shown in these graphs is $\frac{S}{k\tilde V}$.  Also shown are the low temperature and high temperature entropies at low and high densities as computed in equations (\ref{hightld}) and (\ref{hightld1}) and their high density duals.  Figures (\ref{entropy}c) and (\ref{entropy}d) show the same quantities at low and high densities as a function of the dimensionless temperature.  There is good agreement in the region of validity of the approximations.

\begin{figure}[t]
    \centering
    \begin{tabular}{c c}
    (a)&(b)\\
    \includegraphics[width=0.40\textwidth]{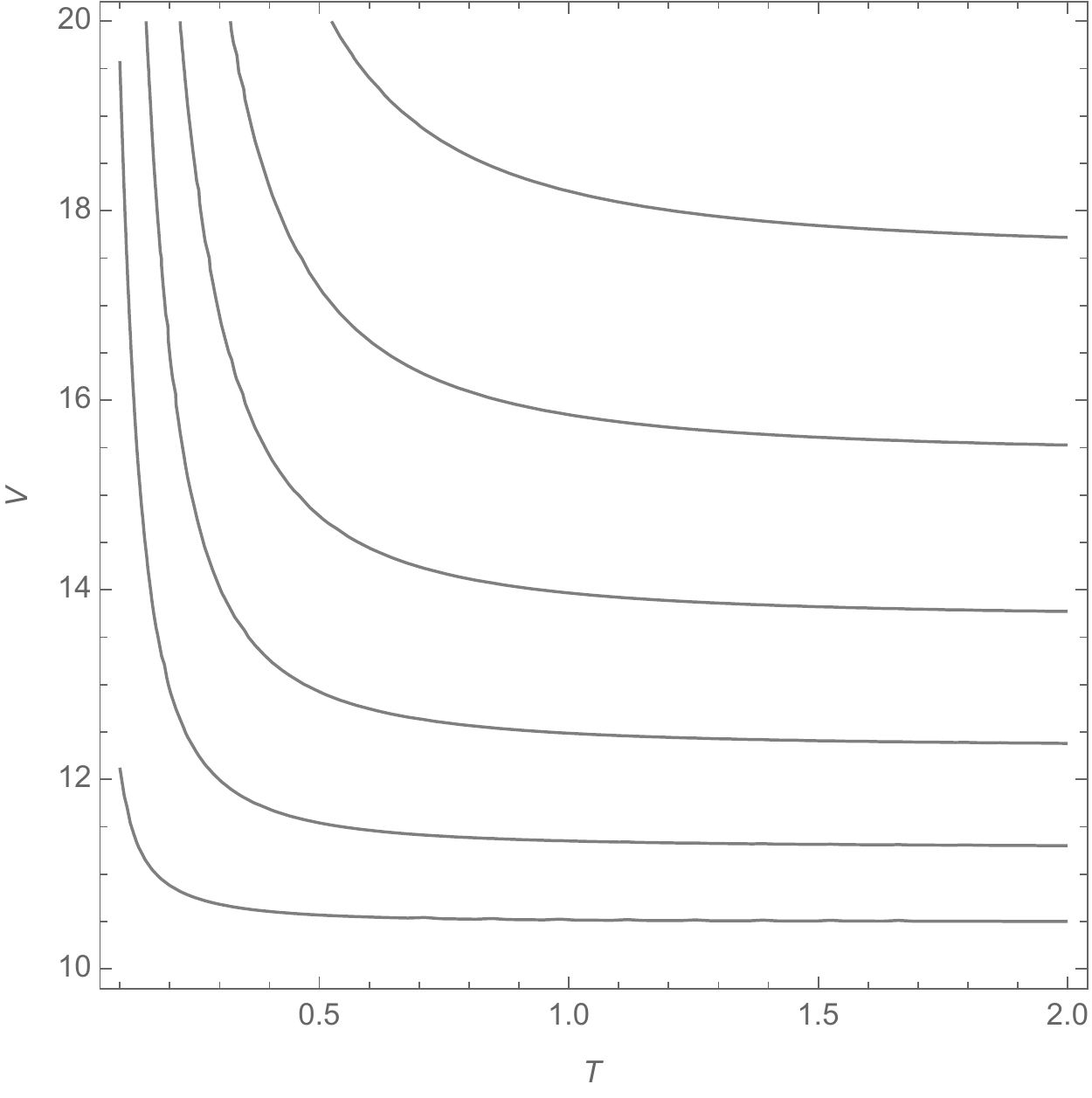}$\quad$&\includegraphics[width=0.40\textwidth]{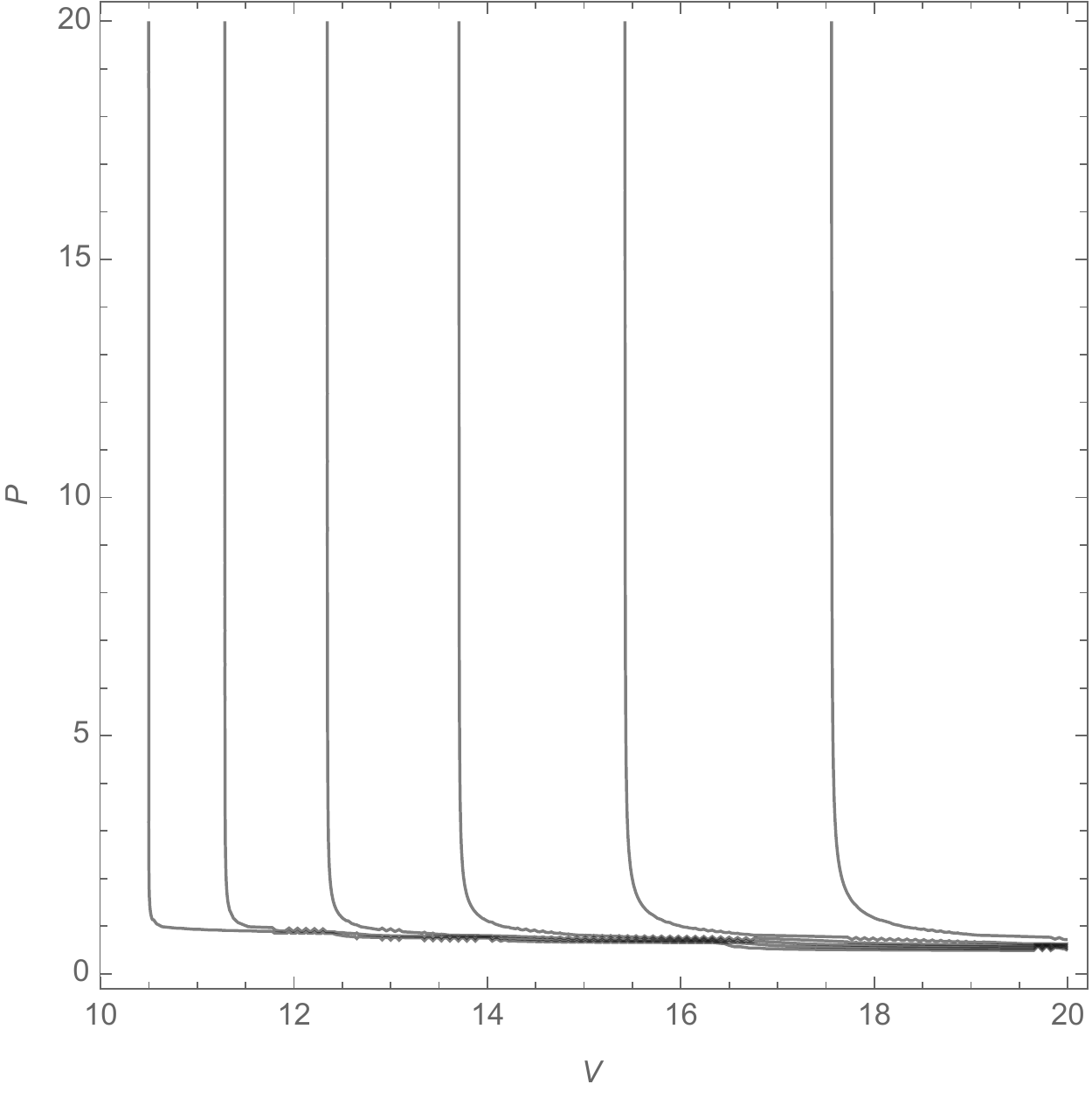}\\
  \end{tabular}
  \caption{Response of the system under adiabatic compression.  Figure (a) shows the dimensionless volume-temperature relationship ($T=\frac{1}{\tilde\beta}$) and (b) the pressure-volume relationship.}
    \label{compress}
\end{figure}

It is useful to also study the response of such a gas under isothermal and adiabatic compression.  Figure (\ref{figpressure}f) shows the isothermals, i..e., the pressure-volume relation when the gas is compressed at fixed temperature.  It exhibits the normal behaviour, except for the existence of the minimal volume where the pressure diverges.  Figure (\ref{compress}a) shows the volume-temperature relation under adiabatic compression. Again the behaviour is standard, except that the volume is again bounded from below by the minimal volume.  Figure (\ref{compress}b) shows the pressure-volume relation under adiabatic compression.  This is the most interesting result as one observes a divergence of the pressure at every value of entropy.  This stems from the fact that entropy decreases with decreasing volume and eventualy vanishes at the minimal volume.  However, if the entropy is fixed, i.e., no heat is allowed to flow into or out of the gas, the system cannot reach volumes with entropy below the initial entropy, which limits the finite volume and manifest itself as a diverging pressure.  

\subsection{The ensemble with $L_3^{\rm tot}\ne 0$.}
The study of the ensemble with $L_3^{\rm tot}\ne 0$ is much more involved as the $\alpha$ and $\lambda$ integrals can no longer be performed explicitly.  However, much progress can be made by realising that the rotating ensemble can be mapped onto an unconstrained ensemble with an effective density of states.  This effective density of states can then be used to analyse the behaviour of these ensembles.  We therefore begin by deriving the equivalent unconstrained ensemble and effective density of states.  The starting point is to rewrite the $q$-potential (\ref{eq:3D-q-pot-cont}) in the following way
\begin{equation}
	 q(M,\tilde\beta,z,\tilde{\omega}) = M^3\int_{-\infty}^{\infty}d\lambda\int_{-\infty}^{\infty}d\alpha\int_{-\infty}^{\infty}dx\, \theta(1-\lambda^2-x^2)\theta(\lambda^2-\alpha^2)\theta(\lambda)\,d(x,\lambda)\log\left[1+ze^{-\tilde{\beta}(x-\tilde\omega\alpha)}\right]
\label{qrewritten}
\end{equation}
This suggests that a transformation from the variables $x,\lambda,\alpha$ to new variables $x^\prime,\lambda^\prime,\alpha^\prime$ can be found that leaves the domain of integration invariant, while transforming away the $\alpha$ dependence in the exponential appearing in the logarithm.  It is simple to see that the desired transformation must satisfy the following conditions:
\begin{eqnarray}
&&x-\tilde{\omega}\alpha=\sqrt{1+\tilde\omega^2}x^\prime,\nonumber\\
&&x^2+\lambda^2={x^\prime}^2+{\lambda^\prime}^2,\nonumber\\
&&\lambda^2-\alpha^2={\lambda^\prime}^2-{\alpha^\prime}^2.
\end{eqnarray}
Solving these conditions yields
\begin{eqnarray}
&&x(x^\prime,\lambda^\prime,\alpha^\prime)=\frac{\tilde\omega\alpha^\prime +x^\prime}{\sqrt{\tilde\omega^2+1}},\nonumber\\
&&\alpha(x^\prime,\lambda^\prime,\alpha^\prime)=\frac{\alpha^\prime-\tilde\omega x^\prime}{\sqrt{\tilde\omega^2+1}},\nonumber\\
&&\lambda(x^\prime,\lambda^\prime,\alpha^\prime)=\frac{\sqrt{\tilde\omega \left(\tilde\omega {x^\prime}^2-2 {\alpha^\prime} {x^\prime}-\tilde\omega{\alpha^\prime}^2 \right)+{\lambda^\prime}^2 \left(\tilde\omega^2+1\right)}}{\sqrt{\tilde\omega^2+1}}.
\end{eqnarray}
Conversely
\begin{eqnarray}
&&x^\prime(x,\lambda,\alpha)=\frac{x- \tilde\omega\alpha}{\sqrt{\tilde\omega^2+1}},\nonumber\\
&&\alpha^\prime(x,\lambda,\alpha)=\frac{\alpha+\tilde\omega x}{\sqrt{\tilde\omega^2+1}},\nonumber\\
&&\lambda^\prime(x,\lambda,\alpha)=\frac{\sqrt{\tilde\omega \left(\tilde\omega x^2+2 \alpha x-\tilde\omega\alpha^2 \right)+{\lambda}^2 \left(\tilde\omega^2+1\right)}}{\sqrt{\tilde\omega^2+1}}.
\end{eqnarray}
One can check the reality of the solution for $\lambda$, $\lambda^\prime$ over the whole integration domain and positivity is ensured by the choice of positive root.  Substituting this transformation into the integral, taking into account the Jacobian, leads to 
\begin{equation}
	 q(M,\tilde\beta,z,\tilde\omega) = M^3\int_{-1}^{1}dx^\prime\int_{0}^{\sqrt{1-{x^\prime}^2}}d\lambda^\prime\int_{-\lambda^\prime}^{\lambda^\prime}d\alpha^\prime\,d^\prime(x^\prime,\lambda^\prime,\alpha^\prime,\tilde\omega)\log\left[1+ze^{-\tilde{\beta}\sqrt{1+\tilde\omega^2}x^\prime}\right]
\label{qtrans}
\end{equation}
where 
\begin{equation}
d^\prime(x^\prime,\lambda^\prime,\alpha^\prime,\tilde\omega)=\frac{\lambda^\prime\sqrt{1-{\lambda^\prime}^2-{x^\prime}^2}}{\pi\left(1-\frac{(\tilde\omega\alpha^\prime+x^\prime)^2}{1+\tilde\omega^2}\right)\sqrt{{\lambda^\prime}^2+{x^\prime}^2-\frac{(\tilde\omega\alpha^\prime+x^\prime)^2}{1+\tilde\omega^2}}}.
\end{equation}
This is exactly the $q$-potential of an unconstrained system ($\tilde\omega=0$) with inverse temperature $\tilde\beta\sqrt{1+\tilde\omega^2}$ if we identify the effective density of states
\begin{equation}
d_{\rm eff}(x^\prime,\tilde\omega)=\int_{0}^{\sqrt{1-{x^\prime}^2}}d\lambda^\prime\int_{-\lambda^\prime}^{\lambda^\prime}d\alpha^\prime\,d^\prime(x^\prime,\lambda^\prime,\alpha^\prime,\tilde\omega),
\end{equation}
although in this case the $\lambda^\prime$ and $\alpha^\prime$ integrals cannot be explicitly computed (actually the $\alpha^\prime$ integral can be performed, but not the $\lambda^\prime$).  It is also useful to note that this can be written as
\begin{equation}
 d_{\rm eff}(x^\prime,\tilde\omega)= \frac{\sqrt{1+\tilde\omega^2}}{\tilde\omega}\int_{0}^{\sqrt{1-{x^\prime}^2}}d\lambda^\prime\int_{u_-}^{u_+}du\,d^\prime(x^\prime,\lambda^\prime,u,\tilde\omega),
\label{qtrans2}
\end{equation}
where $u_\pm=\frac{\pm \tilde\omega\lambda^\prime+x^\prime}{\sqrt{1+\tilde\omega^2}}$ and 
\begin{equation}
d^\prime(x^\prime,\lambda^\prime,u,\tilde\omega)=\frac{\lambda^\prime\sqrt{1-{\lambda^\prime}^2-{x^\prime}^2}}{\pi\left(1-u^2\right)\sqrt{{\lambda^\prime}^2+{x^\prime}^2-u^2}}.
\end{equation}

The thermodynamics following from this $q$-potential can now be studied as before.  For reasons of economy we do not consider all the possible limits, but only the physically most interesting ones of low temperature ($\tilde\beta>>1$) and slowly rotating gases ($|\tilde\omega|<<1$).  In this approximation the $q$-potential reduces to
\begin{equation}
 q(M,\tilde\beta,z,\tilde\omega) =\frac{2M^3}{3\pi}\int_{-1}^1 d{x^\prime}\, \sqrt{1-{x^\prime}^2}\left(\left(1+\frac{\tilde\omega^2(9{x^\prime}^2-2)}{10(1-{x^\prime}^2)} \right)\log\left(1+ze^{-\tilde\beta x^\prime}\right)\right).
\label{lowtloww}
\end{equation}

Before proceeding, we consider the high/low density duality on the level of this approximated $q$-potential.  It is a simple matter to show that this $q$ potential can be rewritten as 
\begin{equation}
 q(M,\tilde\beta,z,\tilde\omega) =\frac{M^3}{3}\left(1+\frac{\tilde\omega^2}{2}\right)\log{z}+q(M,\tilde\beta,z^{-1},\tilde\omega).
\label{lowtlowwdual}
\end{equation}

Let us now consider the low density limit with $\tilde\mu=\frac{1}{\tilde\beta}\log z<-1$.  Then the $q$-potential can be approximated as
\begin{equation}
 q(M,\tilde\beta,z,\tilde\omega) =\frac{2M^3z}{3\pi}\int_{-1}^1 d{x^\prime}\, \sqrt{1-{x^\prime}^2}\left(1+\frac{\tilde\omega^2(9{x^\prime}^2-2)}{10(1-{x^\prime}^2)} \right)e^{-\tilde\beta x^\prime}.
\label{lowtlowwd}
\end{equation}
Performing the integration and restoring dimensions the following results for the thermodynamics to leading order in $L_3^{\rm (tot)}$, $\rho$ and $\beta$
\begin{eqnarray}
&&P=\rho kT,\nonumber\\
&&S=Nk\left[\frac{5}{2}-\log\left(\lambda^3\rho\right)-\frac{325{L_3^{\rm (tot)}}^2 }{784 E_0N^2 m_0 R^2 }\right],\nonumber\\
&&U=NkT\left(\frac{3 }{2 }-\frac{55 {L_3^{\rm (tot)}}^2}{98 E_0N^2m_0R^2}\right)-NE_0+\frac{5{L_3^{\rm (tot)}}^2}{14 Nm_0R^2}.
\end{eqnarray}
It is interesting to note that although the pressure does not exhibit any non-commutative corrections due to the linear approximation in the density, other thermodynamic quantities do exhibit such corrections for the rotating gas.  In this case it is also insightful to compute the energy as the rotational contribution can be identified explicitly.  In this regard also note the shift in the energy by $NE_0$, which simply originates from the shift in energy introduced in (\ref{eq:jacobi-pol-inf-well}).  Further useful physical information can also be extracted from the expression for the dimensionful $\omega$:
\begin{equation}
\omega=\frac{5 L_3^{\rm (tot)}}{7Nm_0R^2},
\end{equation}
from which we identify the moment of inertia as 
\begin{equation}
I=\frac{7Nm_0R^2}{5}.
\end{equation}

Next we consider the high density dual where $\tilde\mu=\frac{1}{\tilde\beta}\log z>1$. The $q$-potential can then be obtained from (\ref{lowtlowwdual}) where the low density $q$-potential is computed as in the case above with $\tilde\mu=\frac{1}{\tilde\beta}\log z<-1$.  One can solve  $\tilde\omega$ and $z$ to leading order in $L_3^{\rm (tot)}$ and $\rho$.  The expression for $z$ is not very informative, but the expression for the dimensional $\omega$ contains useful physical information as the moment of inertia can be identified from it.  For this quantity we find
\begin{equation}
\omega=\frac{3 L_3^{\rm (tot)}}{4\pi \rho_{max} m_0 R^5(1-\frac{1}{\tilde\beta}\log(\lambda^3\rho_h))}.
\end{equation}
From this we identify the moment of inertia as
\begin{equation}
I=\frac{2\pi}{3} \rho_{max} m_0 R^5(1-\frac{1}{\tilde\beta}\log(\lambda^3\rho_h)).
\label{iner}
\end{equation}
Recall that the moment of inertia of a rigid sphere of radius $R$ and matter density $\rho$ is $I=\frac{8\pi}{15}\rho R^5$.  Comparison with (\ref{iner}) suggests that the gas behaves like a body at maximum mass density, yet the deviating factors suggest a form of non-rigidity and/or inhomogeneity.  Note that the logarithmic term can be neglected at very low temperatures and finite hole density.  Interestingly, though, the moment of inertia diverges at fixed temperature and vanishing hole density.
 
For the other thermodynamic quantities we find, after restoring dimensions and keeping only the lowest terms in $\beta$ and $\rho_h$, 
\begin{eqnarray}
 && P=E_0\rho_{max}+kT\rho_h-kT\rho_{max}\log(\lambda^3\rho_h)-\frac{3kT}{8E_0 \pi m_0 N_h R^5(1-\frac{1}{\tilde\beta}\log(\lambda^3\rho_h))^2}{L_3^{\rm (tot)}}^2,\nonumber\\
 && \frac{S}{k}=N_h \left(\frac{5}{2}-\log \left(\lambda ^3 \rho _h\right) \right)+\frac{3\left(3-2 \log \left(\lambda ^3 \rho _h\right)\right)}{16\pi E_0 m_0\rho_{max} 
R^5(1-\frac{1}{\tilde\beta}\log \left(\lambda ^3 \rho _h\right))^2}{L_3^{\rm (tot)}}^2,\nonumber\\
 && U=\frac{3}{2 } N_hkT-N_hE_0-\frac{3}{8\pi m_0\rho_{max} R^5(1-\frac{1}{\tilde\beta}\log \left(\lambda ^3 \rho _h\right))^2}{L_3^{\rm (tot)}}^2.
\end{eqnarray}
Note that, as before, one has to be careful with the order of limits here, i.e., one cannot take the $T\rightarrow 0$ before $\rho_h\rightarrow 0$.

\section{Non-commutative gas confined by gravity}
\label{Section 4}

From the previous sections it is clear that the behaviour of a non-commutative gas at low temperatures and densities does not deviate significantly from that of a commutative gas, which makes any observation of non-commutative effects extremely difficult.  However, at high densities and low temperatures the behaviour of a non-commutative gas deviates significantly from its commutative counterpart, which must have observational consequences in systems where these conditions can be realised.  This may be the case in very dense astro-physical objects, such as white dwarfs and neutron stars, although it is unlikely that the densities of these objects are even high enough for non-commutative effects to manifest themselves.  Despite this, it would still be interesting to study how the non-commutative, non-relativistic equation of state alters the behaviour of a gas confined by gravity.  Of course, one must realise that such a calculation is not realistic in that relativistic effects are important in the study of these systems. Yet, one can expect to gain some insight into the qualitative features of these systems through a non-relativistic calculation.

Motivated by these considerations, we consider here a non-commutative, non-rotating and non-relativistic gas, confined in a large volume, in the presence of gravitational interactions.  One expects that the gas will not fill the entire volume, but that gravity will confine it to a smaller volume.  Our interest will be in the physical properties of this confined gas, particularly the density distribution and the mass-radius relation.  

The starting point of this calculation is, as usual, the condition of hydrostatic equilibrium, which reads in the Newtonian limit, appropriate for a non-relativistic computation,
\begin{eqnarray}
\label{hydrodimf}
&&\frac{dP(r)}{dr}=-\frac{GM(r)\rho_m(r)}{r^2}\nonumber\\
&&\frac{dM(r)}{dr}=4\pi r^2\rho_m(r).
\end{eqnarray}
Here $P(r)$ is the pressure, $M(r)$ the included mass, $\rho_m(r)$ the mass density at radius $r$ and $G$ the gravitational constant.  The mass density is related to the particle density by $\rho_m(r)=m_0\rho(r)$ where $m_0$ is the mass of the constituent particles.  It is useful to first rewrite this in dimensionless form.  Introducing the dimensionless radius $\tilde r=\frac{r}{\theta}$, the dimensionless pressure, $\tilde P$, and particle density, $\tilde\rho$, introduced in section \ref{Section 2}, this can be cast in the form
\begin{eqnarray}
\label{hydrodiml}
&&\frac{d\tilde P(\tilde r)}{d\tilde r}=-\frac{\tilde G\tilde M(\tilde r)\tilde\rho(\tilde r)}{\tilde r^2}\nonumber\\
&&\frac{d\tilde M(\tilde r)}{d\tilde r}= \tilde r^2\tilde\rho(\tilde r).
\end{eqnarray}
Here $\tilde M=\frac{M}{m_0}$ and $\tilde G=\frac{Gm_0^3\theta}{\hbar^2}$ are the dimensionless included mass and gravitational constant.  

If the Fermi energy is large compared to $kT$, which is usually the case in dense astro-physical objects, we can ignore temperature variations and treat the temperature as approximately zero and independent of $\tilde r$.  As the density has a $\tilde r$ dependence, this must be inherited by the dimensionless chemical potential, i.e., $\tilde\mu(\tilde r)$.  From (\ref{eq:3D-q-pot-cont-uncon}) it is simple to see that
\begin{equation}
\label{presg}
\frac{d\tilde P(\tilde r)}{d\tilde r}=\frac{d\tilde \mu(\tilde r)}{d\tilde r}\tilde\rho(\tilde r),
\end{equation}
which implies from (\ref{hydrodiml}) 
\begin{equation}
\tilde M(\tilde r)=-\frac{\tilde r^2}{\tilde G}\frac{d\tilde \mu(\tilde r)}{d\tilde r}.
\end{equation}
From this it is easy to derive an equation for $\tilde\mu(\tilde r)$:
\begin{equation}
\frac{d^2\tilde\mu(\tilde r)}{d\tilde r^2}+\frac{2}{\tilde r}\frac{d\tilde\mu(\tilde r)}{d\tilde r}=-\tilde G\tilde\rho(\tilde r),
\end{equation}
where $\tilde\rho(\tilde r)$ is explicitly given by
\begin{equation}
\tilde\rho(\tilde r)=\frac{2}{\pi}\int_{-1}^1dx\frac{\sqrt{1-x^2}}{1+e^{\tilde\beta(x-\tilde\mu(\tilde r))}}.
\end{equation}
The boundary conditions are
\begin{equation}
\tilde\mu(0)=\tilde\mu_0,\quad \left.\frac{d\tilde\mu(\tilde r)}{d \tilde r}\right |_{\tilde r=0}=0.
\end{equation}
The value of $\tilde\mu_0$ determines the central density and the second condition follows from the requirement of a vanishing pressure gradient at the origin and equation \eqref{presg}. 

\begin{figure}[t]
    \centering
     \includegraphics[width=0.80\textwidth]{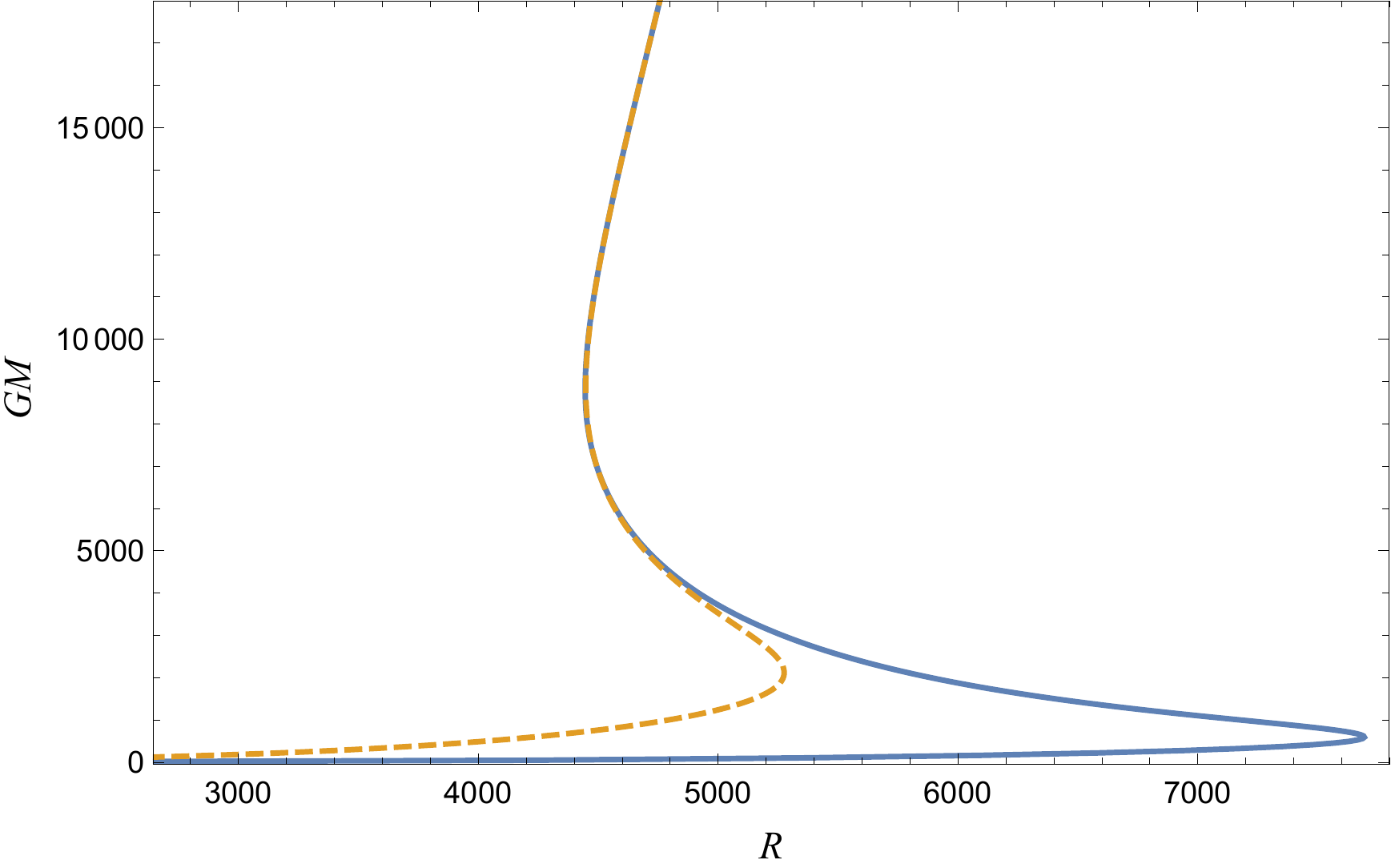}
  \caption{Mass-radius relationship for $\tilde\beta=10$ (dashed line) and $\tilde\beta=50$ (solid line).  For demonstrative purposes the dimensionless gravitational constant was taken as (the unrealistically large) value $\tilde G=10^{-6}$.  All quantities shown are dimensionless and the quantity shown on the vertical axis is $\tilde G\tilde M$.}
    \label{massrad}
\end{figure}

This equation can only be solved numerically and figure \ref{massrad} shows the mass-radius relation for two dimensionless temperatures.  Here the radius has been defined as the point where the dimensionless chemical potential reaches the value $\tilde\mu=-1$ and the density and pressure become exponentially small. Correspondingly the mass is the mass included within this radius.
In contrast to the commutative case, where the hydrostatic equilibrium condition only admits a solution below a threshold mass (the Chandrasekhar mass), there is no threshold mass in the non-commutative case.  This is due to the incompressible behaviour at maximum density, already alluded to in section \ref{Section 3}, when all available single particle states are occupied and the pressure diverges, thereby stabilising the system against collapse.  This can only happen when the dimensionless central chemical potential and pressure exceed the critical value of one and the dimensionless central density saturates at the maximum value of one. In this case one expects the system to behave as an incompressible liquid drop.  

\begin{figure}[t]
    \centering
    \begin{tabular}{c c}
    (a)&(b)\\
    \includegraphics[width=0.40\textwidth]{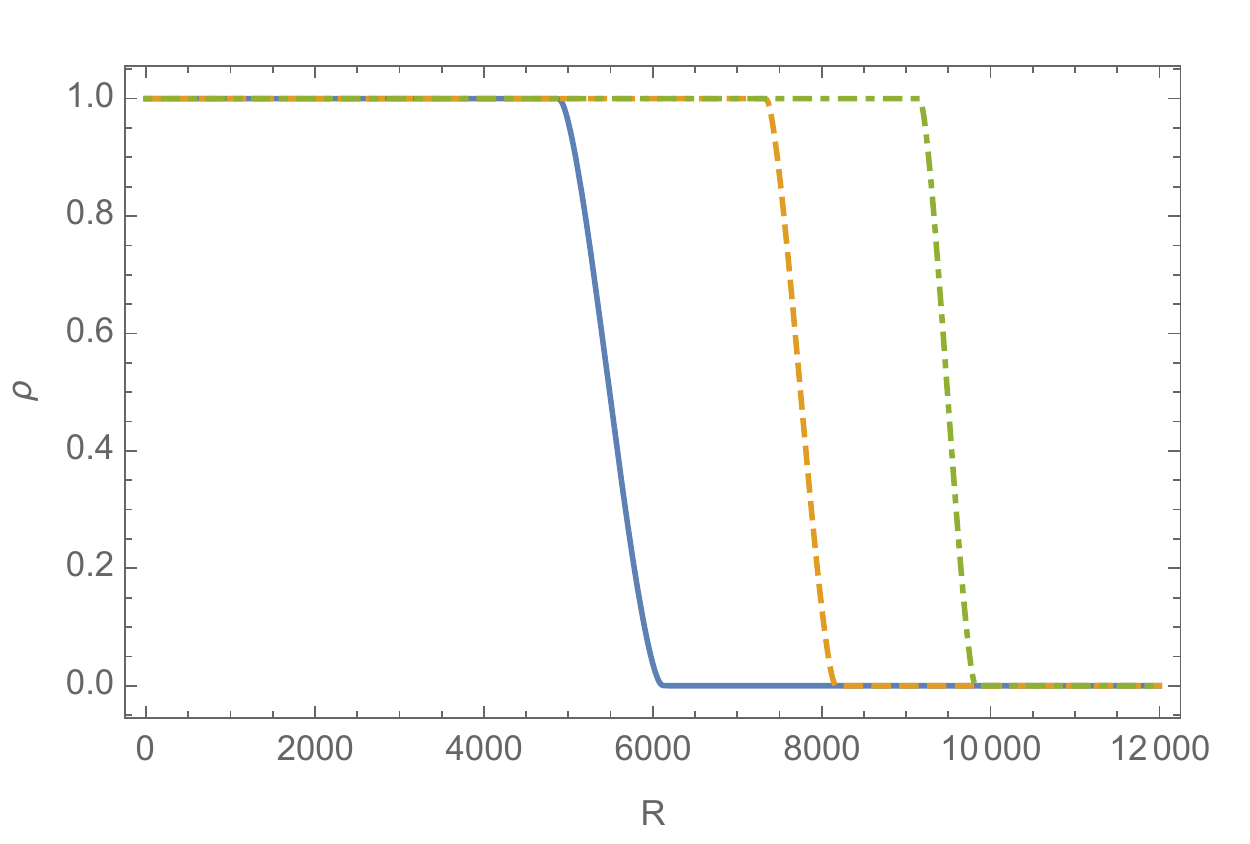}$\quad$&\includegraphics[width=0.40\textwidth]{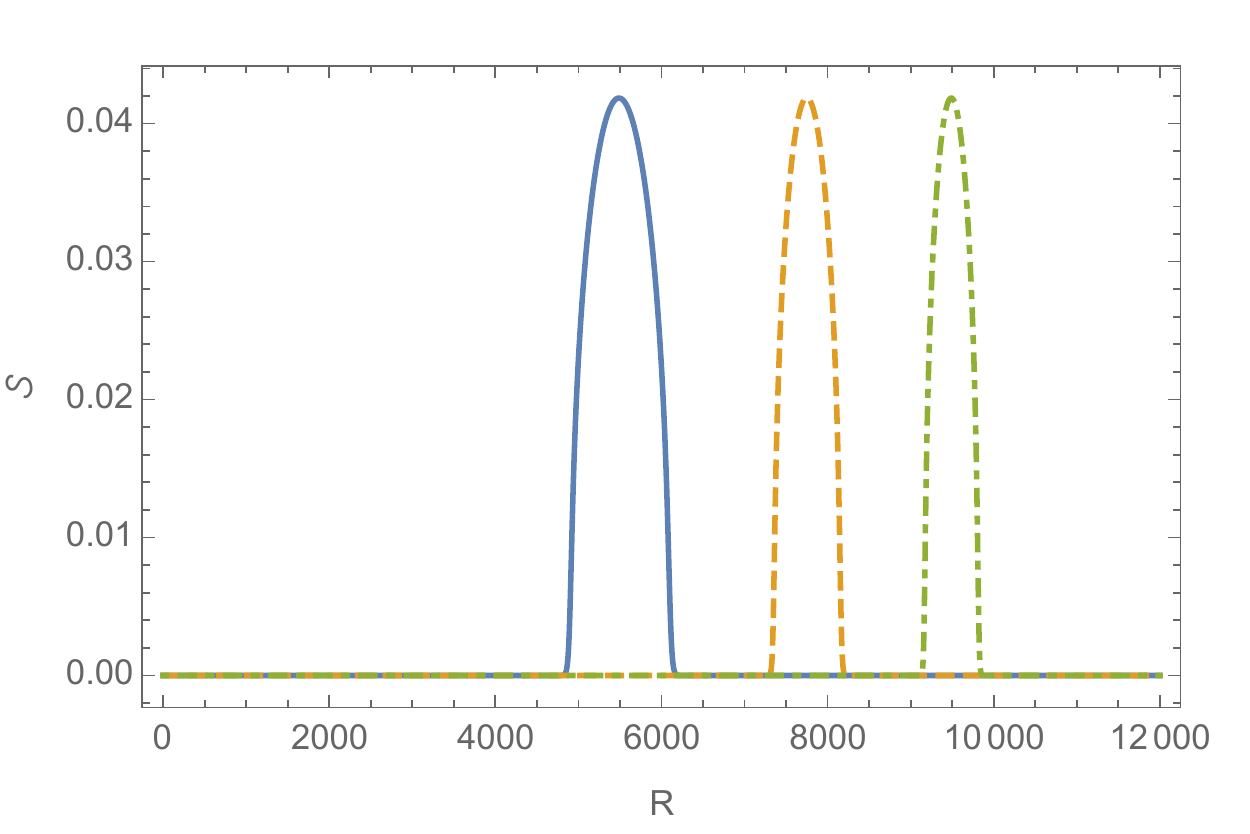}\\
  \end{tabular}
  \caption{(a) Density profile for $\tilde\mu_0=5$ (solid line), $\tilde\mu_0=10$ (dashed line) and $\tilde\mu_0=15$ (dash-dotted line) for $\tilde G=10^{-6}$. (b) Entropy densities for the same parameter sets as in (a).  All quantities shown are dimensionless.}
    \label{density}
\end{figure}

To verify that this is indeed the case we show in figure \ref{density}(a) the density distribution for three values of central chemical potentials larger than one.  From this figure it should be clear that the confined gas behaves as an incompressible liquid drop with maximum density in the bulk and a rapidly vanishing density at the edge.  Indeed, one notes that the edge becomes thinner with increasing radius, indicating that at large mass and radius the incompressible fluid is surrounded by a very thin dilute gas of particles.  This is very reminiscent of quantum Hall liquids at incompressibility, which are known to be closely related to two dimensional non-commutative systems.  The situation encountered here therefore seems like a natural three dimensional generalisation of quantum Hall fluids.  

Due to the incompressible nature  of the bulk, one expects that the excitations of such a fluid will be described by the edge excitations. This is certainly the case for quantum Hall fluids.  One way of verifying this is to define an entropy density as
\begin{equation}
\label{entden}
{\cal S}(\tilde r)=\frac{k}{\tilde V}\left(q(\tilde r)-\tilde\beta\frac{\partial q(\tilde r)}{\partial\tilde\beta}\right),
\end{equation}
where $q(\tilde r)$ is the $q-$potential of (\ref{eq:3D-q-pot-cont-uncon}) evaluated at $\tilde\mu(\tilde r)$.  This is shown in figure \ref{density}(b).  From this it is clear that the entropy vanishes in the bulk and is localised in the thin shell on the surface of the droplet.  This suggests that this is also where the active degrees of freedom of the system must reside, as one would have expected.  

\begin{figure}[t]
    \centering
    \begin{tabular}{c c}
    (a)&(b)\\
    \includegraphics[width=0.40\textwidth]{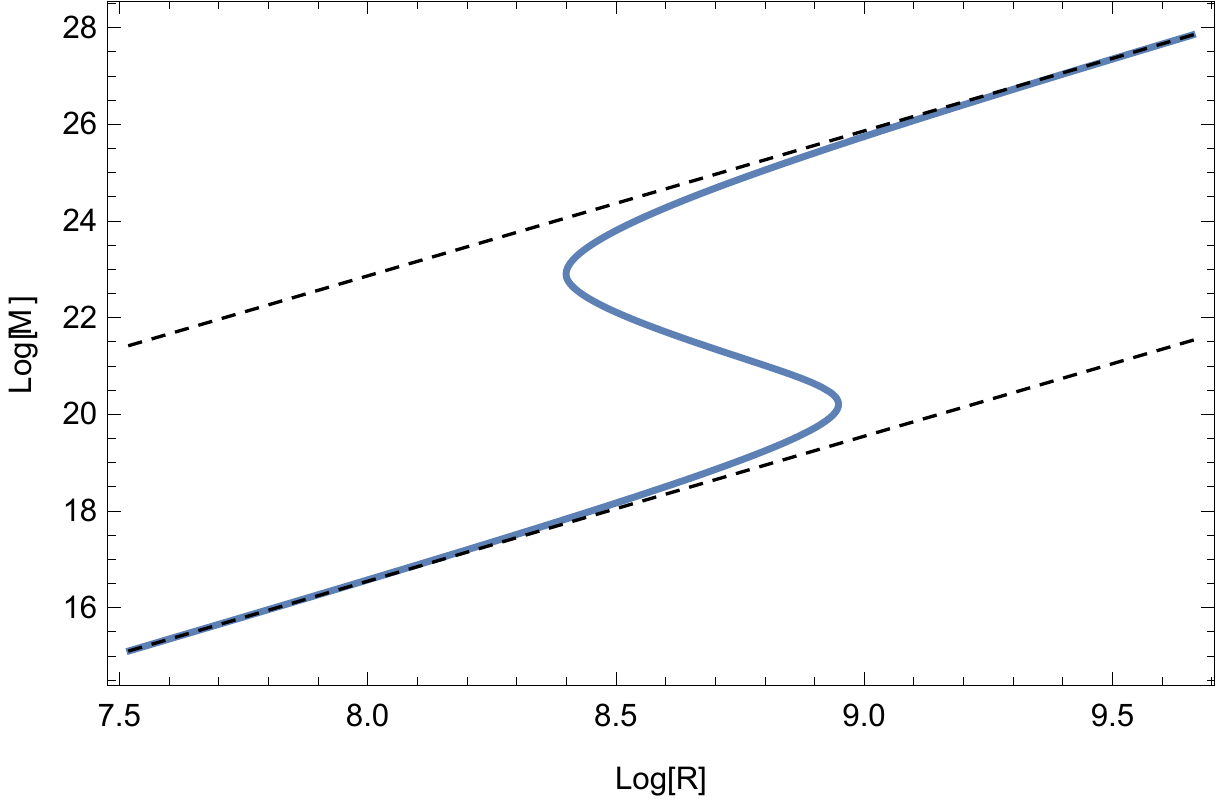}$\quad$&\includegraphics[width=0.40\textwidth]{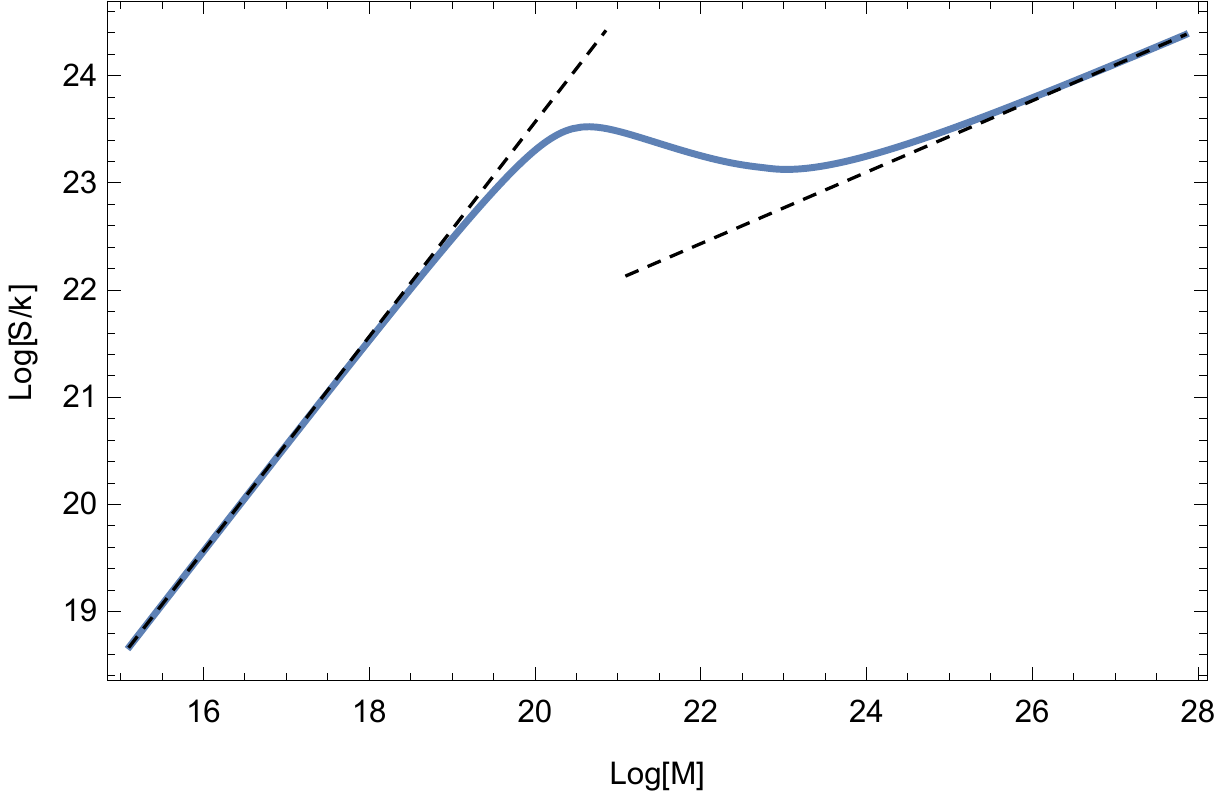}\\
  \end{tabular}
  \caption{(a) Log-log plot of mass against radius for $\tilde\beta=50$.  The dotted lines both have slopes of three.  This demonstrates that the dimensionless mass is extensive ($\tilde M\sim \tilde V$) in both the low and high density regions.  (b) Log-log plot of dimensionless entropy $\tilde S=S/k$ against dimensionless mass.  The left dotted line has a slope of one and the right dotted line of 1/3.  This demonstrates different scaling behaviours of the entropy in the low ($\tilde S\sim \tilde V$) and high  ($\tilde S\sim \tilde V^{1/3}$) density regions with a cross over between the two.}
    \label{evolve}
\end{figure}

The evolution of the system with increasing mass is further clarified in figure \ref{evolve}.  Figure \ref{evolve}(a) shows a log-log plot of dimensionless mass against dimensionless radius.  The two dashed lines both have a slope of three.  This clearly shows that in the low and high density regions the dimensionless mass scales with the dimensionless volume, but with different scaling factors, i.e. densities, as demonstrated by the different intercepts of the two lines on the vertical axis. These two regions correspond to a low density gas and an incompressible liquid drop, as explained above, while there is a cross-over region (the s-curve) in-between.  This cross-over is associated with a change in the scaling behaviour of the entropy, defined as the volume integral of the entropy density (\ref{entden}) over the system volume.  This is demonstrated in figure \ref{evolve}(b), which shows a log-log plot of the dimensionless entropy $\tilde S=S/k$ and dimensionless mass.  The left dotted line has a slope of one and the right one a slope of 1/3.  This implies $\tilde S\sim \tilde V$, i.e. normal extensive behaviour, in the low density region.  In the high density regions, however, the scaling is linear in the radius, i.e $\tilde S\sim \tilde V^{1/3}$ and non-extensive. This is somewhat counterintuitive in that one would have expected a quadratic dependence or area law.  In this regard it should, however, be pointed out that the entropy calculation done here is somewhat naive in two respects: The first is the computation of the entropy as an integral of a rather naively defined entropy density (\ref{entden}).  A more appropriate approach, as used in quantum Hall liquids, would be to derive the effective theory describing the surface excitations of the incompressible fluid and compute the entropy from there.   The second is that the current computation completely ignores the quantised nature of the radius, which should become important in the case of very thin shells with thickness of the order of a few radial quanta.  At this stage it is unclear how these issues should be addressed and they will be pursued elsewhere.

\section{Conclusion}
\label{Section 5}

We have studied the thermodynamics of a non-rotating and slowly rotating gas of fermions confined to a spherical well in three dimensional fuzzy space, with emphasis on non-commutative effects.  These corrections were computed at low temperature and density where the non-relativistic approximation should be perfectly valid. These corrections may provide useful clues in attempts to observe the effects of non-commutativity at low energy scales and low densities.  

As expected non-commutative effects are manifest at high temperatures and densities.  The most prominent of these are the existence of a minimal volume at which the gas becomes incompressible and the fact that the entropy may exhibit non-extensive behaviour at high densities. These systems also exhibit a remarkable low/high density duality, which underpins many of the features observed at high density.  This duality, in turn, is a manifestation of an infra-red/ultra violet duality in the single particle spectrum.

Finally a gas confined by gravity was investigated.  The most remarkable feature that emerged in this case is the formation of an incompressible liquid drop at high central pressure.  This droplet is saturated at maximum density in the bulk with a rapid decay to a low density at the edge, giving rise to a very thin, dilute gas of particles surrounding the droplet.  Entropy calculations suggest that the latter carries the active degrees of freedom of the droplet, which gives rise to a non-extensive behaviour of the entropy as a function of the droplet size.

This study focussed on infinite wells.  Finite wells and other confining potentials offer even more intriguing possibilities.  In particular, finite wells of sufficient depth have a gap between the finite number of bound states and scattering states, which also terminate at high energies \cite{nc-well}.  The scattering states can therefore act as a reservoir for the finite number of bound states. This interplay may have interesting dynamical consequences that will be studied elsewhere.

\end{document}